\documentclass[aps,amssymb,amsmath,pra,twocolumn,showpacs,superscriptaddress,longbibliography]{revtex4-2}

\usepackage{graphicx}% Include figure files
\usepackage{dcolumn}% Align table columns on decimal point
\usepackage{bm}% bold math
\usepackage{subfigure}
\usepackage{color}

\usepackage{amssymb,amsmath,amsfonts,latexsym,graphicx,verbatim}%,dsfont}

\usepackage[margin=0.57in]{geometry}

\usepackage{ulem}

\usepackage{tikz}

\usepackage[english]{babel}
\usepackage{times}
\usepackage{dsfont}
\usepackage{latexsym}
\usepackage{fancyhdr}
\usepackage{float}
\usepackage{afterpage}
\usepackage{enumitem}
\usepackage{mleftright}

\definecolor{lR}{rgb}{1, 0.8, 0.79}

\usepackage{eso-pic, graphicx}

\usepackage{listings}
\usepackage{multirow}
\usepackage{xcolor,colortbl}

\usepackage{bbm}
\usepackage{upgreek}
\usepackage{amsmath}
\usepackage{newtxtext,newtxmath}

\definecolor{Ablue}{rgb}{0.96,0.24,0.00}

\definecolor{Abluetitle}{rgb}{0.,0.24,0.51}

\definecolor{orange}{rgb}{0.96,0.24,0.00}

\definecolor{darkred}{rgb}{0.55, 0.0, 0.0}

\definecolor{darksalmon}{rgb}{0.91, 0.59, 0.48}
\definecolor{maroon}{cmyk}{0,0.87,0.68,0.32}

\setcitestyle{numbers,square,citesep={,\kern-.24em}}

\definecolor{mustard}{rgb}{1.0, 0.86, 0.35}

\addto\captionsenglish{\renewcommand{\figurename}{Fig.}}

\definecolor{Gray}{gray}{0.85}
\definecolor{LightCyan}{rgb}{0.88,1,1}
\newcolumntype{a}{$>${\columncolor{Gray}}c}
\newcolumntype{b}{$>${\columncolor{white}}c}

\usepackage{array}
\newcolumntype{L}[1]{$>${\raggedright\let\newline\\\arraybackslash\hspace{0pt}}m{#1}}
\newcolumntype{C}[1]{$>${\centering\let\newline\\\arraybackslash\hspace{0pt}}m{#1}}
\newcolumntype{R}[1]{$>${\raggedleft\let\newline\\\arraybackslash\hspace{0pt}}m{#1}}

\newcolumntype{P}[1]{>{\centering\arraybackslash}p{#1}}
\newcolumntype{M}[1]{>{\centering\arraybackslash}m{#1}}

\usepackage[colorlinks=true , citecolor=black,urlcolor=blue]{hyperref}

\newcommand{\xa}{\alpha}

\newcommand{\xd}{\delta}

\newcommand{\vxe}{\varepsilon}
\newcommand{\tm}{{\text -}}

\newcommand{\CC}{\R{CC}}

\newcommand{\tacq}{t_{\R{acq}}}
\newcommand{\xg}{\gamma}
\newcommand{\xt}{\vartheta}

\newcommand{\xr}{\rho}
\newcommand{\xo}{\omega}

\newcommand{\app}{\approx}

\newcommand{\Cs}{{}^{13}\R{C}}

\newcommand{\fac}{f_{\R{AC}}}
\newcommand{\fres}{f_{\R{res}}}
\newcommand{\fhet}{f_{\R{het}}}
\newcommand{\Bac}{B_{\R{AC}}}

\newcommand{\wac}{w_{\R{AC}}}
\newcommand{\wl}{w_{\R{L}}}

\newcommand{\mB}[0]{\mathcal{B}}

\newcommand{\mHdd}[0]{\mH_{\R{dd}}}

\newcommand{\kdd}[0]{k_{\R{dd}}}

\newcommand{\xD}{\Delta}

\newcommand{\xO}{\Omega}

\newcommand{\fr}[2]{\frac{#1}{#2}}

\newcommand{\sq}[1]{\sqrt{#1}}

\newcommand{\mH}[0]{\mathcal{H}}

\newcommand{\Hddav}{\overline{\mH}_{\R{dd}}^{(0)}}
\newcommand{\Hzav}{\overline{\mH}_{\R{z}}^{(0)}}

\newcommand{\beq}{\begin{equation}}
\newcommand{\eeq}{\end{equation}}
                  
\newcommand{\benum}{\begin{enumerate}}
\newcommand{\eenum}{\end{enumerate}}
                    
\newcommand{\bit}{\begin{itemize}}
\newcommand{\eit}{\end{itemize}}
\newcommand{\xhat}{\hat{\T{x}}}
\newcommand{\yhat}{\hat{\T{y}}}
\newcommand{\zhat}{\hat{\T{z}}}

\newcommand{\bea}{\begin{eqnarray}}
\newcommand{\eea}{\end{eqnarray}}

\newcommand{\bev}{\begin{verbatim}}
\newcommand{\eev}{\end{verbatim}}

\newcommand{\non}{\nonumber}

\newcommand{\qt}{\tau}

\newcommand{\lb}{\left(}
\newcommand{\rb}{\right)}
\newcommand{\lsb}{\left[}
\newcommand{\rsb}{\right]}

\newcommand{\T}[1]{\textbf{#1}}
\newcommand{\I}[1]{\textit{#1}}
\newcommand{\R}[1]{\textrm{#1}}
\newcommand{\zl}[1]{\label{eqn:#1}}
\newcommand{\zr}[1]{Eq.\,(\ref{eqn:#1})}
\newcommand{\zfl}[1]{\protect\label{fig:#1}}
\newcommand{\zfr}[1]{\figurename\,\ref{fig:#1}}

\newcommand{\zsl}[1]{\label{sec:#1}}
\newcommand{\zsr}[1]{\!\ref{sec:#1}}

\newcommand{\expec}[1]{\left\langle #1\right\rangle}

\newcommand{\ba}{\left\{ \begin{array}{lr}}
\newcommand{\ea}{\end{array}\right.}

\newcommand{\blist}[1]{
 \begin{list}{#1}%$\ast\circ\bullet\Right
 \begin{align}
	 arrow
 \end{align}
 $\checkmark\star
  { \setlength{\itemsep}{3pt}
     \setlength{\parsep}{2pt}
     \setlength{\topsep}{3pt}
     \setlength{\partopsep}{0pt}
     \setlength{\leftmargin}{1em}
     \setlength{\labelwidth}{1em}
     \setlength{\labelsep}{0.5em} } }
\newcommand{\elist}{
  \end{list}  }

\DeclareMathSymbol{\vartheta}{\mathalpha}{letters}{"12}
\DeclareMathSymbol{\theta}{\mathalpha}{letters}{"23}
\DeclareMathSymbol{\phi}{\mathalpha}{letters}{"27}
\DeclareMathSymbol{\varphi}{\mathalpha}{letters}{"1E}

\newcommand{\bef}
{
\begin{figure}[htbp]
\centering
}

\newcommand{\eef}{\end{figure}}

\newcommand*\circled[1]{\tikz[baseline=(char.base)]{
            \node[shape=circle,draw,inner sep=2pt] (char) {#1};}}

\mleftright
\makeatletter
\@addtoreset{section}{part}
\makeatother

\newcommand{\beginsupplement}{%
        \setcounter{table}{0}
        \renewcommand{\thetable}{S\arabic{table}}%
        \setcounter{figure}{0}
        \renewcommand{\thefigure}{S\arabic{figure}}%
				
     }

\newcommand{\affA}{Department of Chemistry, University of California, Berkeley, Berkeley, CA 94720, USA.}
\newcommand{\affB}{Chemical Sciences Division,  Lawrence Berkeley National Laboratory,  Berkeley, CA 94720, USA.}
\newcommand{\affD}{Energy Geoscience Division, Lawrence Berkeley National Laboratory, Berkeley, CA 94720, USA.}
\newcommand{\affE}{OxideMEMS Lab, Purdue University, 47907 West Lafayette, IN, USA.}

\begin{document}
\title{High-Field Magnetometry with Hyperpolarized Nuclear Spins }
\author{Ozgur Sahin}\affiliation{\affA}
\author{Erica de Leon Sanchez}\affiliation{\affA}
\author{Sophie Conti}\affiliation{\affA}
\author{Amala Akkiraju}\affiliation{\affA}
\author{Paul Reshetikhin}\affiliation{\affA}
\author{Emanuel Druga}\affiliation{\affA}
\author{Aakriti Aggarwal}\affiliation{\affA}
\author{Benjamin Gilbert}\affiliation{\affD}
\author{Sunil Bhave}\affiliation{\affE}
\author{Ashok Ajoy}\email{ashokaj@berkeley.edu}\affiliation{\affA}\affiliation{\affB}

\begin{abstract}
Quantum sensors have attracted broad interest in the quest towards sub-micronscale NMR spectroscopy. Such sensors predominantly operate at low magnetic fields. Instead, however, for high resolution spectroscopy, the high-field regime is naturally advantageous because it allows high absolute chemical shift discrimination.  Here we propose and demonstrate a high-field spin magnetometer constructed from an ensemble of hyperpolarized $\Cs$ nuclear spins in diamond. The $\Cs$ nuclei are initialized via Nitrogen Vacancy  (NV) centers and protected along a transverse Bloch sphere axis for minute-long periods. When exposed to a time-varying (AC) magnetic field, they undergo secondary precessions that carry an imprint of its frequency and amplitude.  The method harnesses long rotating frame $\Cs$ sensor lifetimes $T_2^{\prime}{>}$20s,  and their ability to be continuously interrogated. For quantum sensing at 7T and a single crystal sample, we demonstrate spectral resolution better than 100 mHz (corresponding to a frequency precision ${<}$1ppm) and single-shot sensitivity better than 70pT.  We discuss advantages of nuclear spin magnetometers over conventional NV center sensors, including deployability in randomly-oriented diamond particles and in optically scattering media.  Since our technique employs densely-packed $\Cs$ nuclei as sensors, it demonstrates a new approach for magnetometry in the ``coupled-sensor’’ limit. This work points to interesting opportunities for microscale NMR chemical sensors constructed from hyperpolarized nanodiamonds and suggests applications of dynamic nuclear polarization (DNP) in quantum sensing.
\end{abstract}

\maketitle

\begin{figure}[t]
  \centering
  {\includegraphics[width=0.49\textwidth]{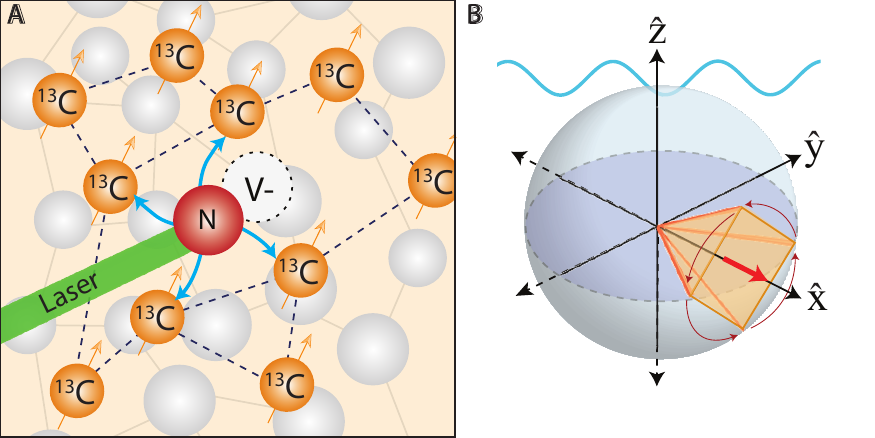}}
  \caption{\T{$\Cs$ Sensor Strategy.} (A) \I{System.} Diamond lattice with dipolar coupled (dashed lines) $\Cs$ nuclei,  hyperpolarized (blue arrows) by optically pumped NV center defects. (B) \I{Principle.} Hyperpolarized $\Cs$ nuclei are driven in to the $\hat{\T{x}}$ axis (red arrow) via spin-locking. When an AC field (blue) is applied along $\hat{\T{z}}$,  the spins undergo secondary precessions (arrows). Deviation from $\hat{\T{x}}$ constitutes the magnetometer signal. Case shown corresponds to $\vartheta{=}\pi/2$ (\zfr{fig2}A), where the spins trace the corners of a square when projected onto the $\hat{\T{y}}\tm\hat{\T{z}}$ plane over four pulses  (period 4$\tau$ in \zfr{fig2}A).}
\zfl{fig1}
\end{figure}

\begin{figure}[t]
  \centering
  {\includegraphics[width=0.49\textwidth]{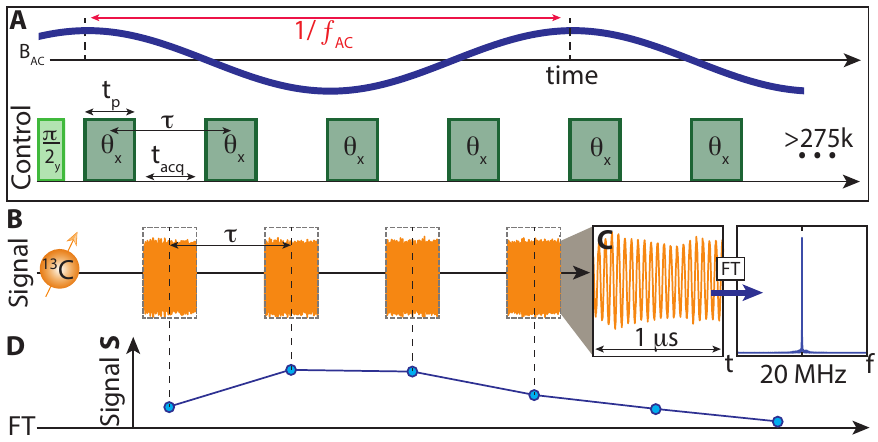}}
  \caption{\T{Magnetometry Protocol.} (A) \I{$\Cs$ sensing sequence} consists of a train of $\xt$ flip-angle pulses (green) separated by interpulse period $\tau$. Upper trace (blue) representatively shows an applied AC field of frequency $\fac$; it is denoted here in the ``resonant” configuration for $\xt{=}\pi/2$. Pulses have width $t_p$, separated by acquisition periods $\tacq$. (B) \I{Directly observed $\Cs$ Larmor precession.} During $\tacq$, $\Cs$ precession is sampled every 1ns. Orange points show representative {raw} data taken in a single-shot for six readout windows between successive pulses.  Here $\tacq{=}32\;\mu$s, $t_p{=}30\;\mu$s, $\qt{=}73\;\mu$s, and $\fac{=}2$ kHz, and phase is not matched with trace in (A).  (C) \I{Zoom into} a 1$\mu$s portion of an acquisition windows.  Fourier transform (blue) displays the heterodyned Larmor precession at $\fhet{=}20$ MHz. (D) \I{Rotating frame signal}. Magnitude of signal at $\fhet$ in each window is plotted (blue points),  extracting the 20 MHz$\pm$32 kHz  component in (C). Line joining points shows oscillations whose frequency components reflects harmonics of $\fac$ (see \zfr{fig5}). }
\zfl{fig2}
\end{figure}

\begin{figure}[t]
  \centering
  {\includegraphics[width=0.49\textwidth]{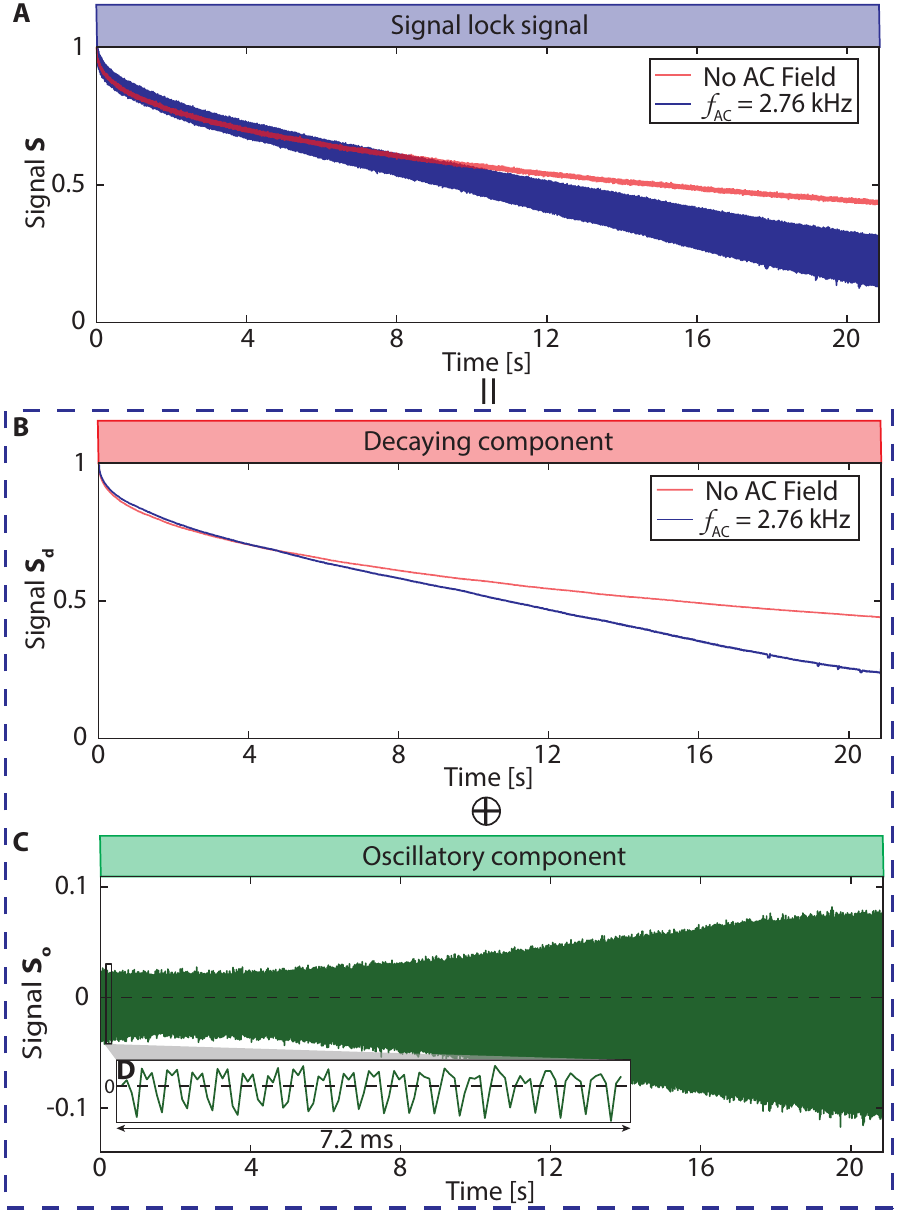}}
  \caption{\T{Long-time $\Cs$ magnetometry signal. } (A) \I{Representative pulsed spin-lock signal} in \zfr{fig2} ($\qt{=}73 \mu s, t_p{=}30\mu$s) with no applied field (red line), and with $\fac{=}$2.76 kHz (blue line), applied near resonance ($\fres{=}2.780$ kHz). The former decays with $T_2’{=}$31 s; AC field, however, yields a faster decay with superimposed oscillations (see \zfr{fig2}A).  (B) \I{Decaying signal component} $S_d$ obtained via a moving average filter. (C) \I{Oscillatory signal component} $S_o$ extracted as $S_0{=}S{-}S_d$. (D) \I{Inset:} zoom into data in a 7.2 ms window (boxed) showing oscillations at $\fac$ (see \zfr{fig5}). }
	\zfl{fig3}
\end{figure}

\section{Introduction}
The discrimination of chemical analytes with sub-micron scale spatial resolution is an important frontier in Nuclear Magnetic Resonance (NMR) spectroscopy~\cite{Sidles09, Meriles05,Degen09}. Quantum sensing methods have attracted attention as a pathway to accomplish these goals~\cite{Degen17}. These are typified by sensors constructed from the Nitrogen Vacancy (NV) defect center in diamond~\cite{Jelezko06,Manson06} --  electrons that can be optically initialized and interrogated~\cite{Gali08,Doherty12}, and made to report on nuclear spins in their environment. However, NV sensors are still primarily restricted to {bulk} crystals and operation at \I{low} magnetic fields ($ B_0{<}$0.3 T)~\cite{Maze08,Boss17,Schmitt17,Glenn18}. Instead, for applications in NMR, high fields are naturally advantageous because the chemical shift dispersion is larger, and analyte nuclei carry higher polarization~\cite{Aslam17}. The challenge of accessing this regime arises from the rapidly scaling electronic magnetogyric ratio $\xg_e$, that makes electronic control difficult at high fields~\cite{Smits19,Fortman20}. Simultaneously, precise field alignment~\cite{Rondin14} is required to obtain viable NV spin-readout contrast~\cite{tetienne12}. The latter has also made nanodiamond (particulate) magnetometers challenging at high fields. If viable,  such sensors could yield avenues for ``\I{targetable}'' NMR detectors that are sensitive to analyte chemical shifts~\cite{Mochalin12,fu07,Nie21},  and provide a sub-micron scale spatial resolution determined by particle size~\cite{Glenn18,Meriles05,Schaffry11}.  In this paper we propose alternate approach towards  overcoming these challenges.  We construct a magnetometer out of hyperpolarized \I{nuclear} spins (See \zfr{fig1}): $\Cs$ nuclei in the diamond serve as the {primary} magnetic field sensors, while NV centers instead play a supporting role in optically initializing them~\cite{Fischer13,Ajoy17}.

The advantages of $\Cs$ nuclei as sensors stem from their attractive properties. Their low $\xg_n{\app}{\xg_e}/3000$, enables control and interrogation at high fields ($B_0{>}$1 T).  In contrast to NV electronic spins, transitions of spin-1/2 $\Cs$ nuclei are determined solely by magnetic field and not influenced by crystal lattice orientation~\cite{Ajoy17,Ajoy20}.  They have long rotating frame lifetimes $T_2'$~\cite{Beatrez21}, orders of magnitude greater than their NV center counterparts.   Similarly,  their longitudinal lifetimes $T_{1}{>}$10 min~\cite{Rej15,Ajoy19} are long even at modest fields. This can allow a physical separation between field regions corresponding to $\Cs$ initialization and sensing, and for the $\Cs$ sensors to be transported between them~\cite{Ajoy19}. $\Cs$ nuclei can be non-destructively readout via RF techniques~\cite{Ajoy20DD}, allowing continuous sensor interrogation without reinitialization. This allows real-time tracking of a changing magnetic field for extended periods~\cite{Shao16}.  RF $\Cs$ readout is also background-free and immune to optical scattering~\cite{lv21, Pillai21},  permitting deployment in real-world media.

While these properties appear attractive at first glance, nuclear spins are often considered ineffective as quantum sensors. The low $\xg_n$, while ideal for high field operation, would be expected to yield low sensitivity~\cite{Degen17}, and  poor state purity (thermal polarization is ${\app} 10^{-5}$ even at 7 T). Strong ($\sim$1 kHz) dipolar coupling between $\Cs$ nuclei makes Ramsey-like sensing protocols untenable. This results in a rapid free induction decay (FID) $T_2^{\ast}{<}$2 ms~\cite{Ajoy20DD},  and limits sensor integration time.

Here, we demonstrate that these shortcomings can be mitigated. We exploit optical hyperpolarization of $\Cs$ nuclei (\zfr{fig1}A)~\cite{Fischer13, London13, Ajoy17,Schwartz18} and spin-lock readout scheme that suppresses  evolution under dipolar interactions~\cite{Beatrez21,Rhim76}. The resulting ${>}$10,000-fold extension in $\Cs$ lifetimes, from $T_2^{*} {\rightarrow} T_2'$, provides the basis for expanding  sensor readout to minute long periods~\cite{Beatrez21}. These long $\Cs$ rotating-frame lifetimes can at least partially offset sensitivity losses arising from the low $\gamma_n$. The sensing strategy is described in \zfr{fig1}B. Hyperpolarized $\Cs$ nuclei are placed along the transverse axis $\hat{\T{x}}$ (red arrow) on the Bloch sphere at high field, where they are preserved for multiple-second long $T_2â€™$ periods~\cite{Beatrez21}. Any subsequent deviation of the spin state from $\hat{\T{x}}\tm\yhat$ plane can be continuously monitored and constitutes the magnetometer signal. In the presence of the target magnetic field, $\T{B}_{\R{ac}}(t){=}\Bac\cos(2\pi\fac t +\phi_0)\hat{\T{z}}$ at a frequency $\fac$, the nuclei undergo a secondary precession in the $\yhat\tm\zhat$ plane that carries an imprint of $\fac$. Long $T_2^{\prime}$ yields high spectral resolution. 

\begin{figure*}[t]
  \centering
  {\includegraphics[width=1\textwidth]{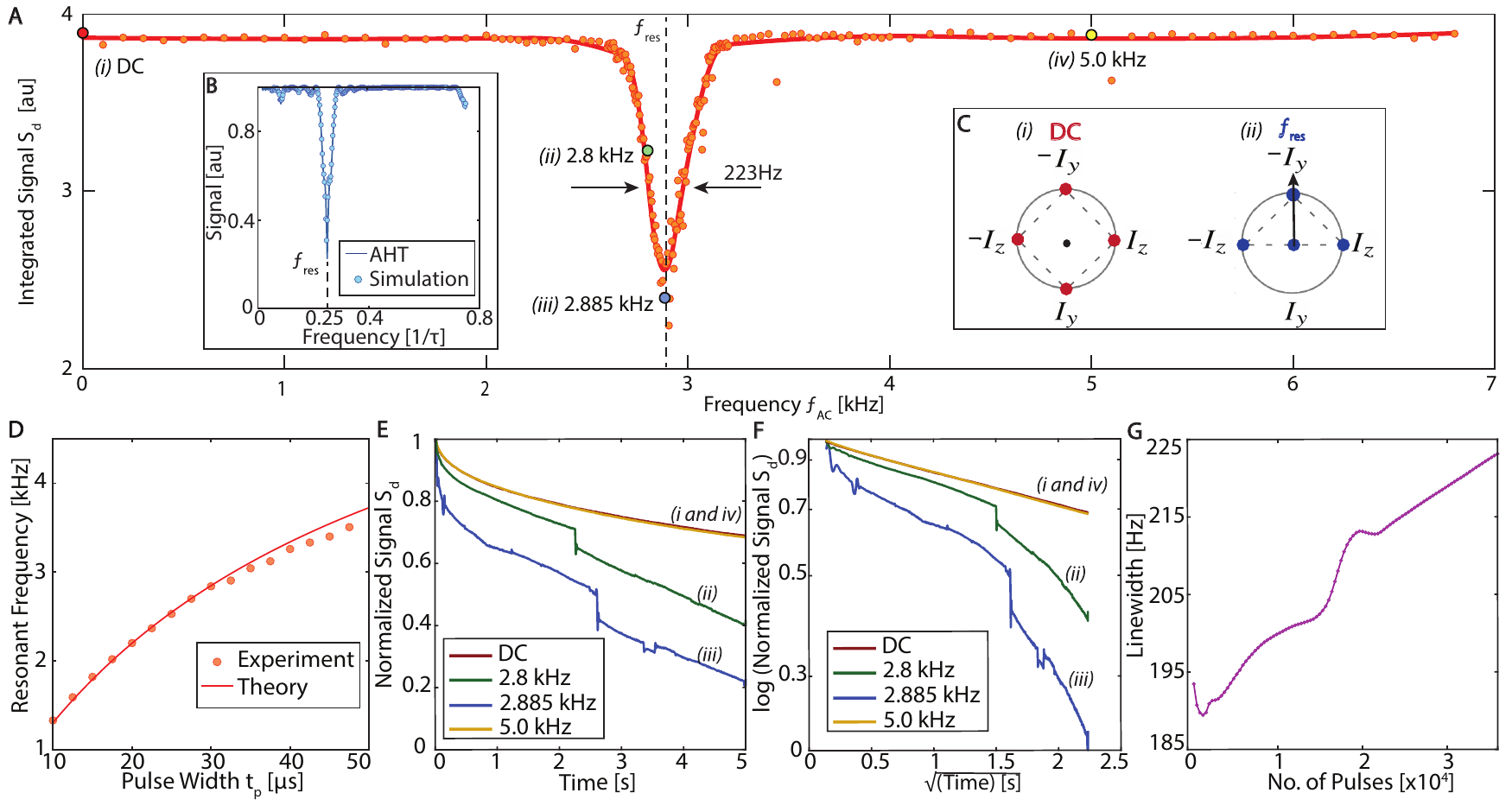}}
  \caption{\T{Enhanced signal decay with applied AC field.} (A) \I{Integrated signal intensity} of  decaying component $S_d$ (see \zfr{fig3}B) with changing AC frequency, with $\xt{=}\pi/2$, and $\tau {=} 73 \mu$s. Decays are normalized against their value at 20 ms and truncated at 5 s (corresponding to ${\app} 10^5$ pulses). Enhanced decay occurs at resonance condition (dashed line). Solid line is a spline fit guide to the eye.  Linewidth is estimated ${\app}$223 Hz (marked) from a Gaussian fit. (B) \I{Average Hamiltonian analysis}. Simulated signal assuming $\xt{=}\pi/2$ and no dipolar coupling between $\Cs$ nuclei. Resonance is expected at $\fres{=}1/4\qt$. (C) \I{Phasor representation} of toggling frame Hamiltonians for (B) under an applied \I{(i)} DC field and \I{(ii)} AC field at resonance.  (D) \I{Resonance frequency scaling}. Points show experimentally extracted resonance frequency from a second harmonic intensity (see \zfr{scale}) for varying pulse widths $t_p$ but fixed $\tacq$. Solid line is a theoretically predicted $\fres$ showing good agreement. (E) \I{Signal traces} for representative points in (A) (large markers \I{i-iv}), corresponding to $\fac{=}$DC, 5kHz, 2.885kHz (resonance) and 2.8kHz (slightly off-resonance). For AC fields near resonance (\I{ii,iii}), we observe rapid decays and sharp jumps (see \cite{SOM}). Far from resonance (\I{i,iv}), the decays exhibit slowly-decaying profiles. (F) \I{Logarithmic scale plot} of the data in (E), plotted against $\sqrt{t}$. Signal decays far from the resonance have a characteristic $\propto\exp(-t^{1/2})$ profile~\cite{Beatrez21}.  Decays close to resonance (\I{ii,iii}), however, show steeper slopes. (G) \I{Scaling of resonance linewidth} in (A) as a function of the number of pulses $N$ applied, estimated from a Gaussian fit.}
\zfl{fig4}
\end{figure*}

\begin{figure*}[t]
  \centering
  {\includegraphics[width=1\textwidth]{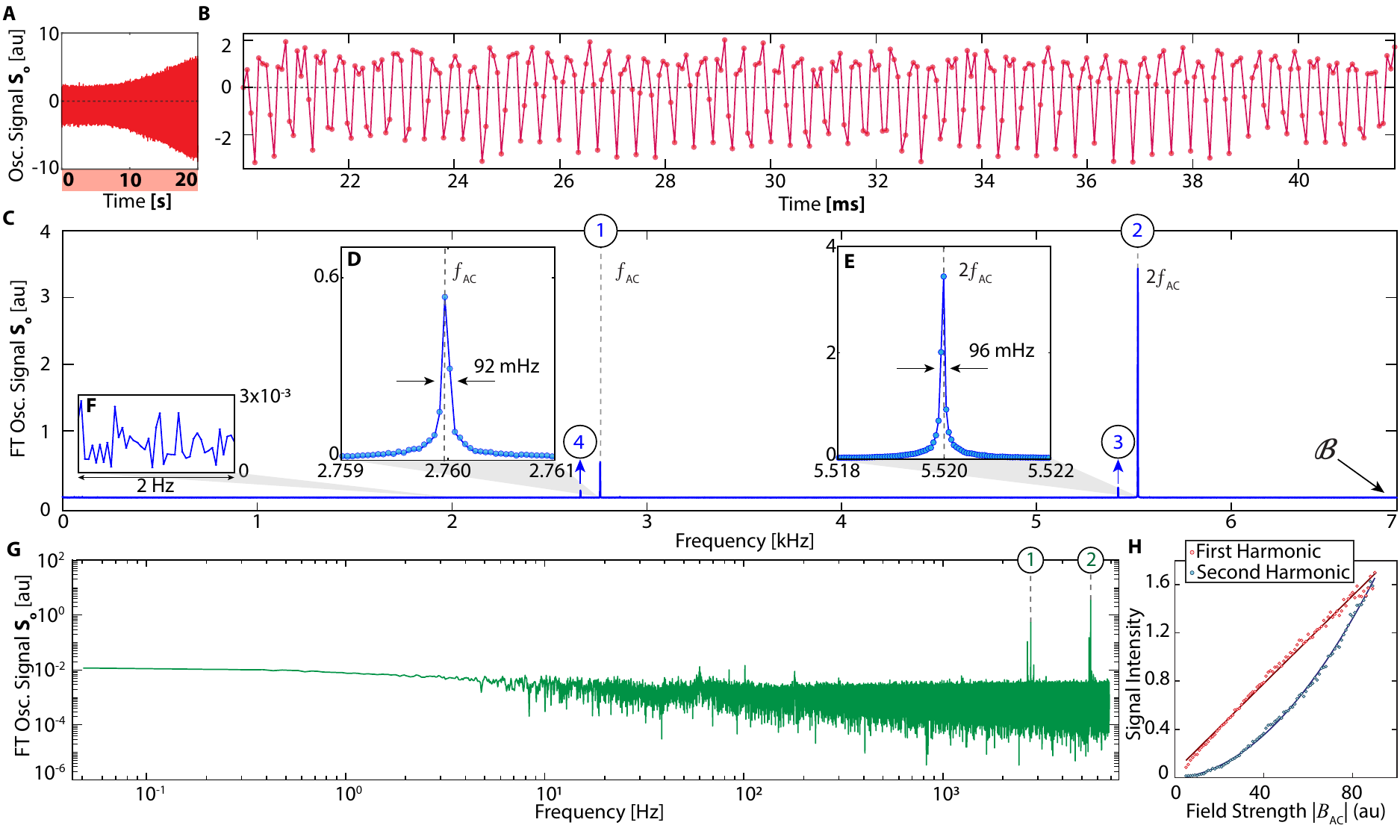}}
  \caption{\T{High-field (7T) $\Cs$ sensor magnetometry.} (A) \I{Long time single-shot signal} of the oscillatory component $S_o$ (following \zfr{fig3}C) upon application of a $\fac{=}$2.760 kHz AC field. Here $\tau {=} 73 \mu$s, $\xt{=}\pi/2$, and $\fres~{=}2.78$ kHz. (B) \I{Zoom into} a 20 ms window. (C) \I{Fourier transform} of data in (A) reveals high resolution peaks corresponding harmonics of $\fac$. Data here is obtained from 100 averages of signal as in (A). Dashed lines represent first and second harmonics. Third and fourth harmonics are aliased, bandwidth $\mB {\app}7$ kHz. (D-E) \I{Zoom into primary and secondary harmonics} centered at $\fac$ and $2\fac$ respectively.  Linewidths are estimated from a Gaussian fit. (F) \I{Zoom into spectral wing} showing noise intensity. (G) \I{Logarithmic plot} corresponding to data in (C). (H) \I{Scaling of harmonic intensity.} Primary and secondary harmonic intensities exhibit apparently linear and quadratic dependence with $|\Bac|$ respectively.}
\zfl{fig5}
\end{figure*}

 \vspace{-3mm}
\section{Results and Discussion} 
 \vspace{-1mm}
As a proof of concept, experiments here are conducted on a single-crystal diamond, but can be extended to powders. The sample has ${\sim}$1 ppm NV center concentration and natural abundance $\Cs$.  Hyperpolarization occurs through a method previously described at 38 mT~\cite{Ajoy17,Ajoy18}.  \zfr{fig2} shows the subsequently applied $\Cs$ magnetometry protocol at 7 T. It entails a train of $\xt$ pulses, spin-locking the nuclear spins along $\hat{\T{x}}$ (\zfr{fig2}A).  $\Cs$ nuclei remain in quasi-equilibrium along $\hat{\T{x}}$ for several seconds~\cite{Beatrez21}.  Flip angle $\xt$ can be arbitrarily chosen, except for $\xt{=}\pi$~\cite{Ajoy20DD}.  Pulse duty cycle is high (19-54\%) and interpulse spacing $\qt{<}100\;\mu$s (\zfr{fig2}A). The nuclei are inductively interrogated in $\tacq$ windows between pulses. Orange points in \zfr{fig2}B show typical raw data.  For each window, the magnitude of the heterodyned Larmor precession, (\zfr{fig2}C, here at 20 MHz) — effectively the rotating frame transverse magnetization component — constitutes the magnetometer signal. These are the blue points in \zfr{fig2}D. Every pulse, therefore, provides one such point; in typical experiments we apply $N{\gtrsim}$200k pulses. In the absence of an external field, successive points show very little decay. However, with an AC field at $\fac$, there is an oscillatory response (evident in blue line \zfr{fig2}D).  It is strongest at the \I{“resonance condition”},
\beq
\fres{=}\fr{\xt}{2\pi\tau} \zl{res}\:,
\eeq
corresponding to when the AC field periodicity is matched to the time to complete a $2\pi$ rotation.  \zfr{fig2}A describes the resonant situation for $\xt{=}\pi/2$.

\zfr{fig3} shows the magnetometer signal over 20 s (275k pulses) with no applied field and $\fac{=}2.760$kHz (close to resonance). In the former case, there is a slow decay $T_2'{=}31$ s (red line in \zfr{fig3}A)~\cite{Beatrez21}.  The AC field however causes a comparatively rapid $\Cs$ decay (blue line in \zfr{fig3}A) along with magnetization oscillations (see \zfr{fig3}C). We separate these contributions, decomposing the signal (dashed box) as $S {=} S_d + S_o$, where $S_d$ is the (slow) decay component (\zfr{fig3}B) and $S_o$ is the oscillatory part with zero mean (\zfr{fig3}C, zoomed in \zfr{fig3}D). In practice, $S_d$ is obtained via a 73 ms moving average filter applied to $S$ in  \zfr{fig2}, and the oscillatory component isolated as $S_o=S-S_d$.

We first focus attention to the decay component $S_d$. In \zfr{fig4}, we vary $\fac$, and study the change in integrated value of $S_d$ over a 5 s period (\zfr{fig3}B).  Here the phase of the applied AC field is random.  \zfr{fig4}A reveals that the $\Cs$ decay is unaffected for a wide range of frequencies, except for a sharp decay response (dip) centered at resonance $\fres$ (dashed vertical line).  With a Gaussian fit, we estimate a linewidth ${\app}223$Hz.  The dip matches intuition developed from average Hamiltonian theory (AHT) (\zfr{fig4}B-C) (see below). It can be considered to be an extension of dynamical decoupling (DD) sensing~\cite{Viola99b,Alvarez11,Witzel07,Bar-Gill12} for arbitrary $\xt$. \zfr{fig4}D shows the scaling of $\fres$ with pulse width $t_p$,  and agrees well with theory,  assuming $t_p{=}36\mu$s for $\xt{=}\pi/2$.

\zfr{fig4}E-F, display individual decays in \zfr{fig4}A on a linear scale and logarithmic scale against $\sqrt{t}$. Points far from resonance (e.g. \I{(i)} DC and \I{(iv)} 5 kHz) exhibit a stretched exponential decay $\propto\exp(-t^{1/2})$, characteristic of interactions with the P1 center spin bath~\cite{Beatrez21}. These manifest as the straight lines in \zfr{fig4}F. On the other hand, for points within the $\fres$ dip in \zfr{fig4}A, we observe a potentially chaotic response evidenced by sharp jumps (\zfr{fig4}E-F). These features belie a simple explanation from AHT alone. The jumps occur for strong Rabi frequencies, and when $\fac$ approaches resonance, evident in the comparison between $\fac{=}5$kHz and $\fac{=}2.885$kHz in \zfr{fig4}E-F (blue and yellow lines) (see SI\ ~\zsr{sec5}).  Elucidating the route to chaos here is beyond the scope of this manuscript and will be considered elsewhere.  \zfr{fig4}G describes the linewidth dependence of the obtained resonance dip as a function of the number of pulses employed. Contrary to DD sensing~\cite{Ajoy2017, Boss17,Gefen19,Rotem19}, the linewidth does not fall with increasing number of pulses, suggesting it is dominated by $\Cs$ dipolar couplings. 

Despite this relatively broad linewidth, high resolution magnetometry can be extracted from the oscillatory component $S_o$ (\zfr{fig3}C). \zfr{fig5}A shows $S_o$ over 20 s for $\fac{=}2.760$ kHz. \zfr{fig5}B zooms into a representative 22 ms window. Strong $\Cs$ oscillations are evident here. Taking a Fourier transform, we observe four sharp peaks (\zfr{fig5}C). We identify the two strongest peaks as being \I{exactly} at $\fac$ and $2\fac$ (shown in \zfr{fig5}C with the labels $\circled1$ and $\circled2$) ; we will refer to them as primary and secondary harmonics respectively. They are zoomed for clarity in \zfr{fig5}D-E (along with the noise level in \zfr{fig5}F), from where we extract the AC magnetometry linewidths as 92 mHz and 96 mHz respectively. Two other smaller peaks in \zfr{fig5}C are aliased versions of the third and fourth harmonics (marked $\circled3-\circled4$), at frequencies $f_3=2\mB-3\fac =5.418$ kHz and $f_4=2\mB-4\fac =2.658$ kHz.  The bandwidth $\mB{=}1/(2\qt)$ here is determined by the interpulse interval in \zfr{fig2}A. For clarity, \zfr{fig5}G shows the data in \zfr{fig5}C in a logarithmic scale, with the harmonics marked. 

We emphasize differences with respect to DD quantum sensing~\cite{Boss17, Schmitt17}.  Oscillations here are at the \I{absolute} AC frequency $\fac$,  as opposed to the difference frequency from $\qt^{-1}$ in the DD case. Second, sensing can be obtained with an arbitrary flip angle $\xt$ (except $\pi$), allowing for greater robustness compared to DD sequences that require precisely calibrated $\xt{=}\pi$ pulses. Pulse error affects AC magnetometry peak intensity but not their position. \zfr{fig5}H shows the scaling of the harmonic intensities with $|\Bac|$. We observe a linear and quadratic dependence of the primary and secondary harmonic intensities (see \zfr{fig5}H), differing from DD sensing. 

We now perform experiments to determine the \I{frequency response} of the sensor, unraveling the sensitivity profile at different frequencies (\zfr{fig6}).  We seek to determine how it relates to the $\fres$ dip in \zfr{fig4}.  We apply a chirped AC field, $\fac(t){=} f_{\R{ini}} + (\xD\mB/T)t$, with $f_{\R{ini}}{=}1$ kHz, in a 1-4 kHz window ($\xD\mB{=}3$kHz) in \zfr{fig6}A (shaded region.  The sweep is slow, ($T{=}20$ s), and occurs only once during the full sequence, and does not start synchronously with it.  If the frequency response was independent of frequency, we would expect an approximately box-like Fourier signal intensity over $\xD\mB$. Instead, we obtain a narrow response with a central cusp in a small portion of the $\xD\mB$ band (zoomed in \zfr{fig6}B).  This response is strongest close to resonance (\zr{res}). Simultaneously, we observe a strong Gaussian response in the region outside $\xD\mB$ (\zfr{fig6}C). We identify these features as the primary and secondary harmonic response respectively (labeled $\circled1$ and $\circled2$ as before). Indeed, their frequency centroids are in the ratio 1:2 (see \zfr{fig5}C); $\fres$ arises where the cusp dips to its lowest point (dashed line in \zfr{fig6}B). We hypothesize that the cusp arises because the actual signal response reflects a dispersive lineshape, flipping in sign on either side of resonance $\fres$ (see Sec.  \zsr{theory}).

\zfr{fig6} suggests the potential for real-time magnetic field \I{``tracking”}. We consider this experimentally in \zfr{fig7}. \zfr{fig7}A shows $S_o$ corresponding to a single-shot of the experiment in \zfr{fig6}A. Sensor response is enhanced as the frequency crosses resonance $\fres$ (dashed line). Three 15 ms windows are displayed in \zfr{fig7}B. Their Fourier transforms reveal that the frequency response follows the instantaneous AC field $\fac(t)$. 

\begin{figure}[t]
  \centering
  {\includegraphics[width=0.49\textwidth]{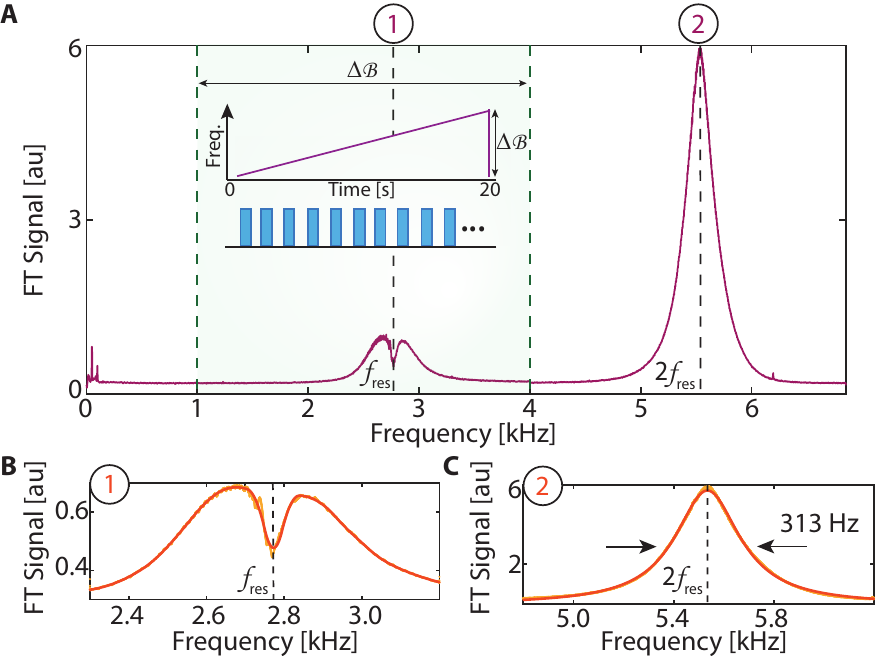}}
  \caption{\T{Frequency response of $\Cs$ magnetometer.} (A) \I{Spectral response} of an chirped AC field (\I{inset}) where frequency is swept between 1-4 kHz ($\xD\mB{=}3$kHz), (green shaded region) in 20 s. Plotted is the resulting Fourier transform intensity of $S_o$ (averaged over 100 shots) revealing profile of sensor frequency response. Data is smoothed over 10 Hz for clarity. Primary and secondary harmonic responses are marked. Dashed lines denote $\fres$ and $2\fres$.  (B) \I{Zoom into primary harmonic response}. Points show cusp-like primary harmonic response with a linewidth of ${\app}60$ Hz about $\fres {=}2.78$ kHz (dashed line). Orange solid line is a spline fit guide to the eye. (C) \I{Zoom into secondary harmonic response}showing an approximately Gaussian profile, here with a linewidth of ${\app}313$ Hz, centered at 2$\fres$.}
\zfl{fig6}
\end{figure}

 \vspace{-3mm}
\section{Theory} 
\zsl{theory}
 \vspace{-1mm}
We now turn to a theoretical description of $\Cs$ sensor operation. We outline two complementary viewpoints to explain the observations in \zfr{fig4} and \zfr{fig5}: a first picture equivalent to DD quantum sensing, and second using an ``rotating-frame’’ NMR experiment analogue. Consider first that the $\Cs$ Hamiltonian in the lab-frame is, $\mH=\mH_{\R{Z}}+\mH_{\R{dd}}+\mH_{\R{AC}}$, where $\mH_{\R{Z}}{=}w_\R{L}I_z$ is the Zeeman Hamiltonian, $\mH_{\R{dd}}{=}\sum_{k<\ell}b_{k\ell}(3I_{kz}I_{\ell z}-\vec{I_k}\cdot \vec{I_\ell})$ is the interspin dipolar interaction and $\mH_{\R{AC}}{=}\gamma_n B_{\R{AC}}\cos(2\pi f_\R{{AC}}t+\phi_0)I_z$ is the applied AC field. Here $I$ refer to spin-1/2 Pauli matrices, $w_\R{L}$ is the nuclear Larmor frequency, $\phi_0$ is the initial (arbitrary) phase of the AC field, and we estimate the median dipolar coupling $J{=}\expec{b_{k\ell}}{\app}$663Hz. The spins are prepared initially along $\hat{\T{x}}$ in a state $\xr(0){\sim} \vxe I_x$, where $\vxe{\app}0.2\%$ is the hyperpolarization level. 

The sequence in \zfr{fig2}A can be conveniently treated in the rotating frame by average Hamiltonian theory (AHT)~\cite{Haeberlen76}. After $N$ pulses, its action can be described by the unitary, $U(N\tau){=}\lsb\exp(i\xt I_x)\exp(i \mH \qt)\rsb^N$, where we assume $\xd$-pulses for simplicity. The signal obtained following the procedure in \zfr{fig2} then simply corresponds to the measurement of $[\expec{I_x}^2 + \expec{I_y}^2]^{1/2}$, or equivalently the spin survival probability in the $\xhat\tm\yhat$ plane. This evolution can be expressed as,  $U(N\qt){=}\prod_{j=1}^{N}\exp(i\mH^{(j)}\qt)$, where $\mH^{(j)}$ are toggling frame Hamiltonians after every pulse~\cite{Haeberlen76}, $\mH^{(j)}{=} \exp(ij\xt I_x) \mH  \exp(-ij\xt I_x)$. For time $t{=}N\qt$, this can be recast as, $U(t){=}\exp(i\mH_F N\qt)$, where $\mH_F$ captures the effective system dynamics under the pulses.  To leading order in parameter $\zeta{=}2\pi J\qt$ in a Magnus expansion~\cite{Magnus54,Wilcox67,Blanes09}, and assuming $\zeta{\ll}1$, the dynamics can be captured by an average Hamiltonian, ${\mH}^{(0)}_{F}= \sum_{j=1}^{N}\mH^{(j)}$. 

Since non-commuting effects do not affect the leading order term, we can consider the effect of each of the dipolar part and the AC field separately, ${\mH}^{(0)}_{F}{=}{\mH}^{(0)}_{\R{dd}} + {\mH}^{(0)}_{\R{AC}}$. For the former, we have~\cite{Ajoy20DD},
\beq
{\mH}^{(0)}_{\R{dd}}= \sum_{j=1}^{N}\mH^{(j)}_{\R{dd}} \app \sum_{j<k}d_{jk}^{\CC}\lb \fr{3}{2}\mH_{\R{ff}}-\vec{I_j}\cdot\vec{I_k}\rb,
\zl{H0}
\eeq 
with the flip-flop Hamiltonian, $\mH_{\R{ff}} = I_{jz}I_{kz}+ I_{jy}I_{ky}$. The initial state $\xr(0)$ is conserved under ${\mH}^{(0)}_{\R{dd}}$, since $[\xr(0){,}{\mH}^{(0)}_{\R{dd}}]{=}0$. This leads to a quasi-equilibriation of spins along $\xhat$, with a lifetime that scales with a power law of the pulsing frequency $\qt^{-1}$~\cite{Beatrez21}. This is the red signal in \zfr{fig3}A (here $\zeta{=}$0.19). We note that for sufficiently small $\zeta$, \zr{H0} is valid for arbitrary flip-angle $\xt$, except for certain special values ($\xt{\app} \pi, 2\pi$). 

A similar AHT analysis can be carried out for the $\mH_{\R{AC}}$ term. Consider first a DC field ($f_{\R{AC}}{=}0$) and $\vartheta{=}\pi/2$. \zfr{fig2}C shows the toggling frame Hamiltonians  $\mH^{(j)}_{\R{AC}}$, which consists only of single body terms and hence can be plotted in a phasor representation. In a cycle consisting of four pulses (required to complete a 2$\pi$ rotation), the average Hamiltonian ${\mH}^{(0)}_{\R{AC}}{=}0$, evident from the symmetrically distributed phasors in \zfr{fig4}\I{(i)}. Hence the DC field is decoupled. Alternately, consider the resonant AC case ($f_{\R{AC}}{=}\fres$). The analysis here is simplest to carry out assuming a square-wave (as opposed to sinusoidal) field. In this situation, the phasor diagram is asymmetrical and the average Hamiltonian after four pulses, ${\mH}^{(0)}_{\R{AC}}{\propto}-I_y$. When stroboscopically observed, the spins rotate away from $\xhat\tm\yhat$ plane; this yields the dip in the integrated signal in \zfr{fig4}.

An illuminating alternate viewpoint is obtained by noting that under spin-locking the spins are requantized in the rotating frame with an effective field, $\xO_{\R{eff}}=\xO\lb \fr{t_p}{\qt}\rb$, where the factor in brackets is the pulsing duty cycle,  and $\Omega$ is the Rabi frequency.  The resonance frequency identified above is then exactly, $\fres{=}\xO_{\R{eff}}$. Therefore the experiment can cast as an rotating-frame \I{analogue} of a conventional (lab-frame) NMR experiment: the spins are quantized along $\xhat$, $\xO_{\R{eff}}$ serves analogous to the Larmor frequency, and at the resonance condition ($\fac{=}\fres$), $\Bac$ is the effective Rabi frequency. Spins initially prepared along $\xhat$, are then constantly tipped away from this axis by $\fac$. The dip in \zfr{fig4} reflects this tilt away from the $\xhat\tm\yhat$ plane. For the resonant case, the trajectory of the spins in the rotating frame can be  simply written as,
\bea
\xr(t) &=& I_x\cos(\xg_n \Bac t) + I_y\sin(\xg_n \Bac t)\sin(2\pi\fac t) \non\\
&+& I_z\sin(\xg_n \Bac t)\cos(2\pi \fac t) \zl{NMR}\:.
\eea
In effect, the spins are undergoing a ``secondary’’ precession in the rotating frame around $\xhat$ at frequency $\xO_{\R{eff}}$. For each point on this motion, they are also precessing in the lab frame at $w_\R{L}$. The latter yields the inductive signal measured in the raw data in \zfr{fig2}B. Upon taking a Fourier transform (\zfr{fig2}C), we are able to extract the magnitude of the spin vector in the $\xhat\tm\yhat$ plane, which has the form, 
$
S(t){=} \lsb \cos^2(\xg_n \Bac t) + \sin^2(\xg_n \Bac t)\sin^2(2\pi \fac t)\rsb^{1/2}
$.  This is the signal measured in \zfr{fig3}, and the oscillations here at $\fac$ and $2\fac$, corresponding to the observed first and second harmonics respectively.  While this analysis was for the resonant case, it is simple to extend it to off-resonant AC fields (see SI\: \zsr{off-res}). Once again the oscillations can be demonstrated to be at exact harmonics of $\fac$.  Alternatively, if the FT phase (instead of magnitude) was taken in \zfr{fig2}C,  where one measures the phase of the spins in the $\xhat\tm\yhat$ plane in the rotating frame, \zr{NMR} indicates there will once again be oscillations.  Here however, all the intensity will be in the primary harmonic, and for sufficiently small fields $\xg_n|\Bac|{\ll}\fac$, the oscillatory signal scales ${\propto}|\Bac|$, and can provide higher sensitivity.  Extracting these oscillations comes with complications related to unwrapping the phase every  $2\pi$, and accounting for phase accrual during the $t_p$ pulse periods. We will detail an approach to perform this phase unwrapping in follow up work. 
 
Overall \zr{NMR} illustrates that the oscillations observed in \zfr{fig5} are equivalent to observing the Larmor precession of the spins in the rotating frame. Since the sequence suppresses static $\zhat$-field inhomogeneity, the lifetime of the oscillations can extend up to $T_2^{\prime}$ (as evidenced in \zfr{fig5}A). The growing oscillation strength in \zfr{fig5}A indicates the spins tipping further away from the $\xhat$ axis.  However, projections of the spin vector away from $\xhat$ are susceptible to dipolar evolution, and the true amplitude of the oscillations observed depend on an interplay between $T_2^{*}$ and $\fac$. We will consider this in detail in follow up theoretical work. Finally, we note that \zr{NMR} demonstrates that \zfr{fig6}-\zfr{fig7} is analogous to the AC field driving a rapid adiabatic passage in the rotating frame.  Therefore, the sign of $\expec{I_x}$ flips on either side of resonance (see \zfr{fig6}). 

\begin{figure}[t]
  \centering
  {\includegraphics[width=0.49\textwidth]{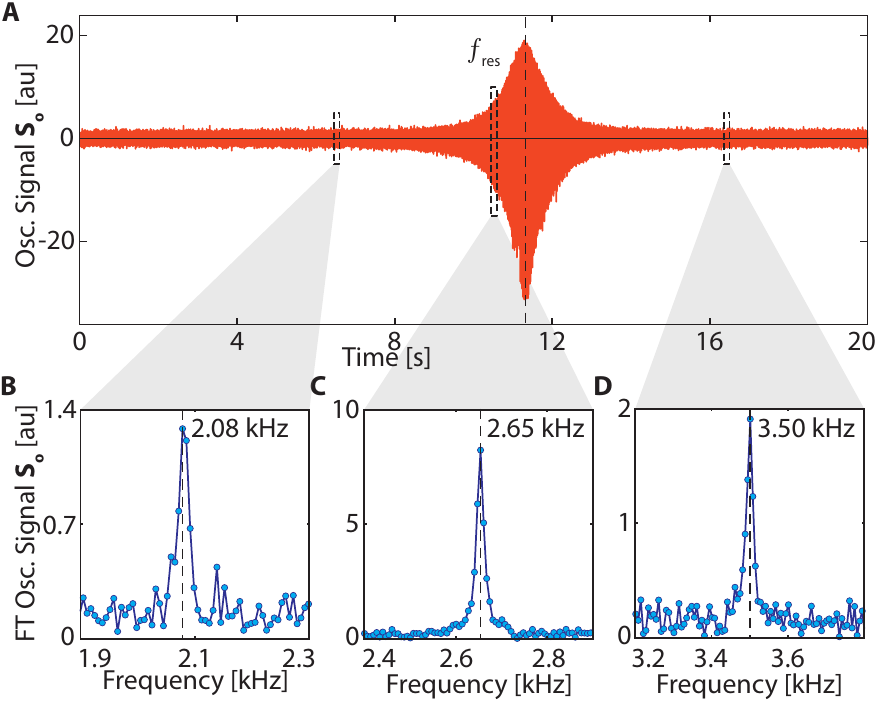}}
  \caption{\T{Tracking signals in time domain} at 7T using a $\Cs$ magnetometer. (A) \I{Long-time oscillatory response} $S_o$ under the chirped AC field in \zfr{fig6}A. Signal shows a strong response as the instantaneous frequency cross resonance $\fres$. (B-D) \I{Tracking of AC frequency}. Three 150 ms time windows are marked, and panels (B-D) show the corresponding Fourier transforms. Solid lines are Lorentzian fits. Clearly the $\Cs$ signal carries a real-time imprint of the frequency of the applied chirped AC magnetic field. }
\zfl{fig7}
\end{figure}

 \vspace{-3mm}
\section{Discussion of $\Cs$ sensor performance and special features}
 \vspace{-1mm}
We now elucidate parameters of $\Cs$ sensor performance and discuss some of its special features (see SI~\cite{SOM} for a summary). Our emphasis in this paper was not to optimize sensitivity,  but through experiments similar to \zfr{fig4} (detailed in SI\: \zsr{sens}) for a bias field $ B_0{=}7$ T, we obtain a single-shot sensitivity of $760\pm 127$ pT$/\sqrt{\R{Hz}}$ via measurement of signal amplitude, and $410\pm90$ pT$/\sqrt{\R{Hz}}$ via measurement of signal phase respectively.  This neglects hyperpolarization time because a sensor initialization event can allow interrogation for several minutes~\cite{Beatrez21}.  A single-shot of the measurement (for $t{=}34$s) can detect a minimum field of  ${\app}$70 pT.  While the sensitivity here is lower than NV quantum sensors at low field,  we emphasize that this corresponds to an exquisite AC field precision of ${\sim}10^{-11}$ over the 7 T bias field.  As such,  the sensitivity here is competitive with respect to other high-field sensors (see SI\: \zsr{sec11} for a detailed comparison),  while also allowing advantages in the multiplexed sensing of AC fields that is not feasible via other sensor technologies at high field (SI\: \zsr{sec11}).

Sensitivity itself can be significantly enhanced. Sample filling factor (presently $\eta{\app}$10\%) and the hyperpolarization level (${\app}$ 0.3\%) can both be boosted by close to an order of magnitude. Employing a 10\% $\Cs$ enriched sample would provide a further ten-fold gain in inductive signal strength~\cite{Ajoy20DD}. From this, we estimate a sensitivity approaching 2 nT$/\sqrt{\text{Hz}}$ is feasible in a (10$\mu$m)$^3$ volume, sufficient to measure fields from precessing ${}^{1}$H nuclei in the same volume at high field~\cite{Glenn18}.

The frequency resolution of the $\Cs$ sensor is $\xd f{\app}{1}/{N\tau}$. Currently, \zfr{fig5} demonstrates a resolution better than 100mHz. However, finite memory limitations restricted capturing the $\Cs$ Larmor precession here to $t{<}35$s (\zfr{fig5}A). Overcoming these memory limits can allow acquisition of the entire spin-lock decay, lasting over 573 s~\cite{Beatrez21}.  Under these conditions, we estimate a frequency resolution of 2.2 mHz is feasible. This would correspond to a frequency precision of 3 ppt at a 7 T bias field, more than sufficient precision to discriminate chemical shifts.  On the other hand, sensor bandwidth $\mB{=}1/2 \tau$ is determined by the minimum interpulse delay (see \zfr{fig4}C). Current Rabi frequencies limit bandwidth to ${\sim}$20 kHz. We estimate that improving filling-factor $\eta$ and RF coil Q factor (presently 50), could increase $\mB$ further to ${\app}500$ kHz. The sensor strategy lends itself to a wide operating field range. While our experiments were carried out at 7 T, the slow scaling $\Cs$ magnetogyric ratio makes sensing viable even for fields  ${\gtrsim}$24 T. This greatly expands the field range for spin sensors, where the operating field is predominantly ${<}$0.3 T (notable exceptions are Refs.~\cite{Aslam17,Fortman20}, but require complex instrumentation).

We emphasize the robustness of our sensing method to pulse error. It can be operated with any flip angle $\xt{\neq}\pi$~\cite{Ajoy20DD}. Moreover, $\fres$ has a relatively wide profile ${\app}$230 Hz (see \zfr{fig4}A), meaning that flip-angle (RF) inhomogeneity has little impact. These features are responsible for the ${>}275$k pulses applied to the $\Cs$ spins in experiments in this work.

Finally, we note some special features of our magnetometry protocol. Compared to previous work~\cite{Granwehr05} using hyperpolarized gaseous ${}^{129}$Xe nuclei as sensors, our experiments employed hyperpolarized nuclear spins in solids. This provides natural advantages due to an ability for in-situ replenishment of hyperpolarization at the sensing site. Moreover, multiple AC fields can be discerned in a Fourier reconstruction in a  single-shot, as opposed to point-by-point~\cite{Granwehr05}.

Fundamentally,  experiments here illustrate the feasibility of quantum sensing in the \I{coupled sensor} limit~\cite{Zhou20,Dooley18}, making the spins sensitive to external fields while negating the effect of intersensor interactions. Sensor operation exploits ``Floquet prethermalization''~\cite{Else16, Khemani16, Howell19, Ueda20} -- quasi-equilibrium nuclear states under periodic driving~\cite{Beatrez21}. As such, this provides a compelling demonstration of exploiting {stable} non-equilibrium phases~\cite{Else16,Khemani16,Luitz20} for sensing applications. 

 From a technological perspective, RF interrogated sensing, as described here, presents advantages in scattering environments~\cite{lv21} (see SI\: \zsr{sec10}). All data in this paper are carried out with the diamond immersed in ${\sim}$4mL water, over 2000-fold the volume of the sample.  Traditional NV sensors are ineffective in this regime due to scattering losses and concomitant fluorescence fluctuations.  Similarly, optically hyperpolarized sensors present advantages because majority of the sensor volume can be illuminated by the impinging lasers, because there are no geometrical constraints from requirements of collection optics. In our experiments, we employed an array of low-cost laser diode sources~\cite{Sarkar21} for hyperpolarization that illuminates the sample almost isotropically. This allows recruiting a larger volume of spins for sensing with a low cost overhead~\cite{Sarkar21}. Extension to powder samples could be advantageous for optimally packing a sensor volume.  

 \vspace{-5mm}
\section{Outlook}
 \vspace{-1mm}
The work presented here can be extended in several promising directions. First, the high resolution-to-bandwidth ratio ($\xd f/\mB {\app} 10^6$) possible via $\Cs$ sensing at high fields (see \zfr{fig5}), suggests possibilities for detecting chemical shifts from precessing analyte nuclei external to the diamond.  Seminal work by Warren et al. ~\cite{Warren93,Richter95} and Bowtell~\cite{Bowtell92} showed that nuclear spins of one species (``sensor'') could be used to indirectly probe NMR information of other physically separated (``analyte'') nuclei. This indirect NMR detection strategy is appealing as the absolute dimensions diminish to sub-micron length scales~\cite{Meriles05}. Building on these ideas, we envision the possibility of micro-scale NMR detectors using hyperpolarized $\Cs$ nuclei in nanodiamond (ND) particles. These experiments require control applied to the target nuclei in order to heterodyne chemical shifts to within the $\Cs$ sensor bandwidth~\cite{Mamin13,Meinel21}, and only one-half of the particle surfaces to be exposed to the analyte~\cite{Warren93}. We will explore these experiments in a future manuscript.

Second, while the current experiments exploit a low-field DNP mechanism for the initialization of the $\Cs$ sensors,  interesting opportunities arise from employing complementary \I{all-optical} DNP techniques that operates directly at high field~\cite{King10,Scott16}. This will allow in-situ sensors at high field without the need for sample shuttling.  

More broadly,  this work suggests an interesting applications of DNP for quantum sensing. It is intriguing to consider sensor platforms constructed in optically-active molecular systems~\cite{Bayliss20,Yu21} where abundant, long-lived nuclear spins, can be initialized via interactions with electronic spin centers. Finally, we envision technological applications of the $\Cs$ sensors described here for bulk magnetometry~\cite{Lesage12}, sensors underwater and in scattering media, and for spin gyroscopes~\cite{Ajoy12g,Ledbetter12,Jaskula19,Jarmola21}.

 \vspace{-3mm}
\section{Conclusions}
 \vspace{-1mm}
We have proposed and demonstrated a high-field magnetometry approach with hyperpolarized $\Cs$ nuclear spins in diamond. Sensing leveraged long transverse spin $\Cs$ lifetimes and their ability to be continuously interrogated, while mitigating effects due to interspin interaction. We demonstrated magnetometry with high frequency resolution (${<}$100mHz), high field precision (${\sim}10^{-11}$),  and high-field (7T) operation,  yielding advantages over counterpart NV sensors in this regime. This work opens avenues for NMR sensors at high fields, and opens new and interesting possibilities for employing dynamic nuclear polarization for quantum sensing. 

We gratefully acknowledge M. Markham (Element6) for the diamond sample used in this work,  and discussions with J. Reimer, C. Meriles, D.Suter and H. Oaks-Leaf. This work was partially funded by ONR under N00014-20-1-2806.  Support to B.G. was provided by U.S.  DOE Office of Basic Energy Sciences, Chemical Sciences, Geosciences, and Biosciences Division under Contract DE-AC02-05CH11231. 

\bibliography{13C_sensing20.bbl}
\pagebreak

\clearpage
\onecolumngrid
%\begin{widetext}
\begin{center}
\textbf{\large{\textit{Supplementary Information} \\ \smallskip
High field magnetometry with hyperpolarized nuclear spins}}\\
\hfill \break
\smallskip
O. Sahin$^{1}$, E.  Sanchez$^{1}$, S.  Conti$^{1}$, A.  Akkiraju$^{1}$, P. Reshetikhin$^{1}$, 
E. Druga,$^{1}$  A. Aggarwal,$^{1}$ B. Gilbert$^{2}$, S. Bhave$^{3}$ and A. Ajoy$^{1,4}$\\
${}^{1}$\I{{\small Department of Chemistry, University of California, Berkeley, Berkeley, CA 94720, USA.}}\\
${}^{2}$\I{{\small Energy Geoscience Division, Lawrence Berkeley National Laboratory, Berkeley, CA 94720, USA.}}\\
${}^{3}$\I{{\small OxideMEMS Lab, Purdue University, 47907 West Lafayette, IN, USA.}}\\
${}^{4}$\I{{\small Chemical Sciences Division Lawrence Berkeley National Laboratory,  Berkeley, CA 94720, USA.}}\\
\end{center}

\twocolumngrid
\beginsupplement
\tableofcontents
\begin{figure}[t]
  \centering
  {\includegraphics[width=0.5\textwidth]{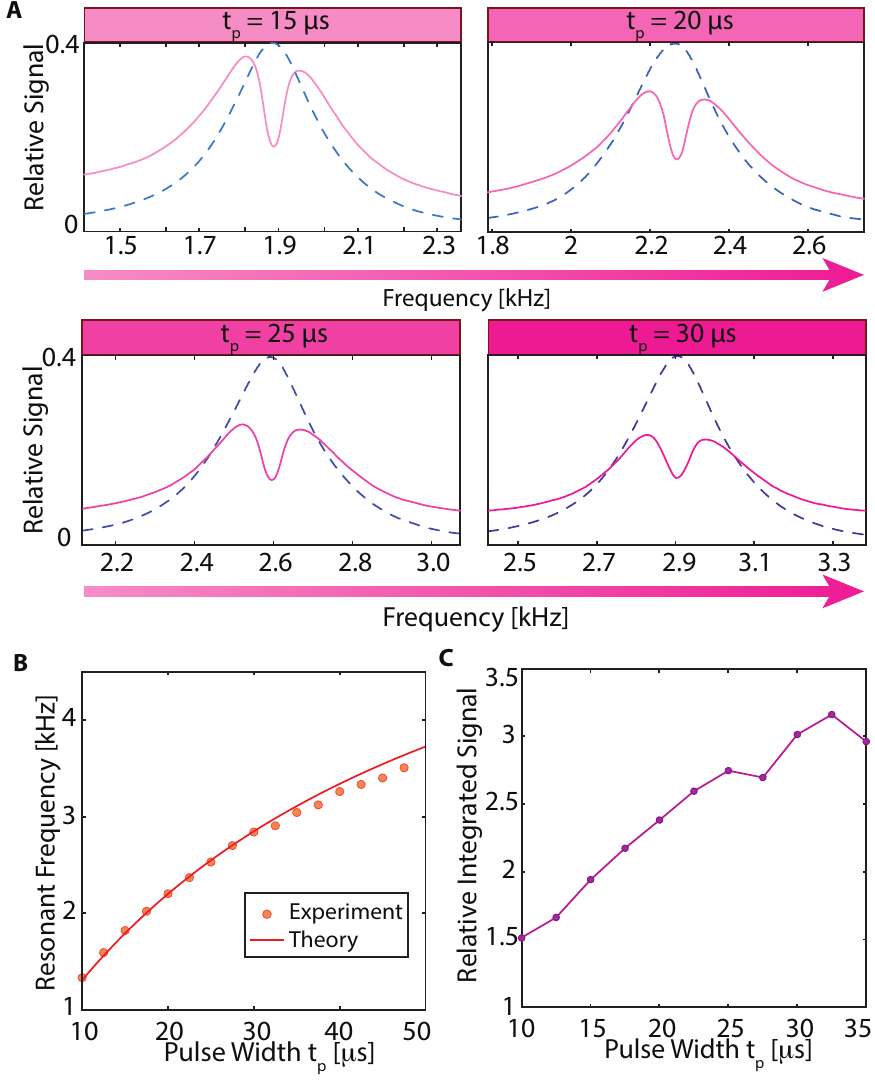}}
  \caption{\T{Scaling of primary and secondary harmonic intensities. } (A) Zoomed in windows corresponding to the primary (pink line) and secondary (dashed blue line) harmonic frequency response for pulsed spin-lock protocol with varying pulse duty cycle.  Traces are normalized to second harmonic peak. While the frequency response remains identical, relative strength of the primary harmonic with respect to secondary harmonic decreases with increasing pulse duty cycle (arrow). (B) Scaling of the $\Cs$ magnetometer resonance frequency with respect to interpulse spacing, estimated experimentally by halving the second harmonic peak frequency. Lines show theoretically predicted resonance frequencies at different applied flip angles.  There is a good agreement with the theoretical prediction \zr{res} (solid line). (C) Relative magnitude of the secondary harmonic intensity to the primary harmonic intensity, obtained from (A). Data indicates a redistribution of signal power to the second harmonic with increasing pulse width.}
\zfl{scale}
\end{figure}

\begin{figure}[t]
  \centering
  {\includegraphics[width=0.49\textwidth]{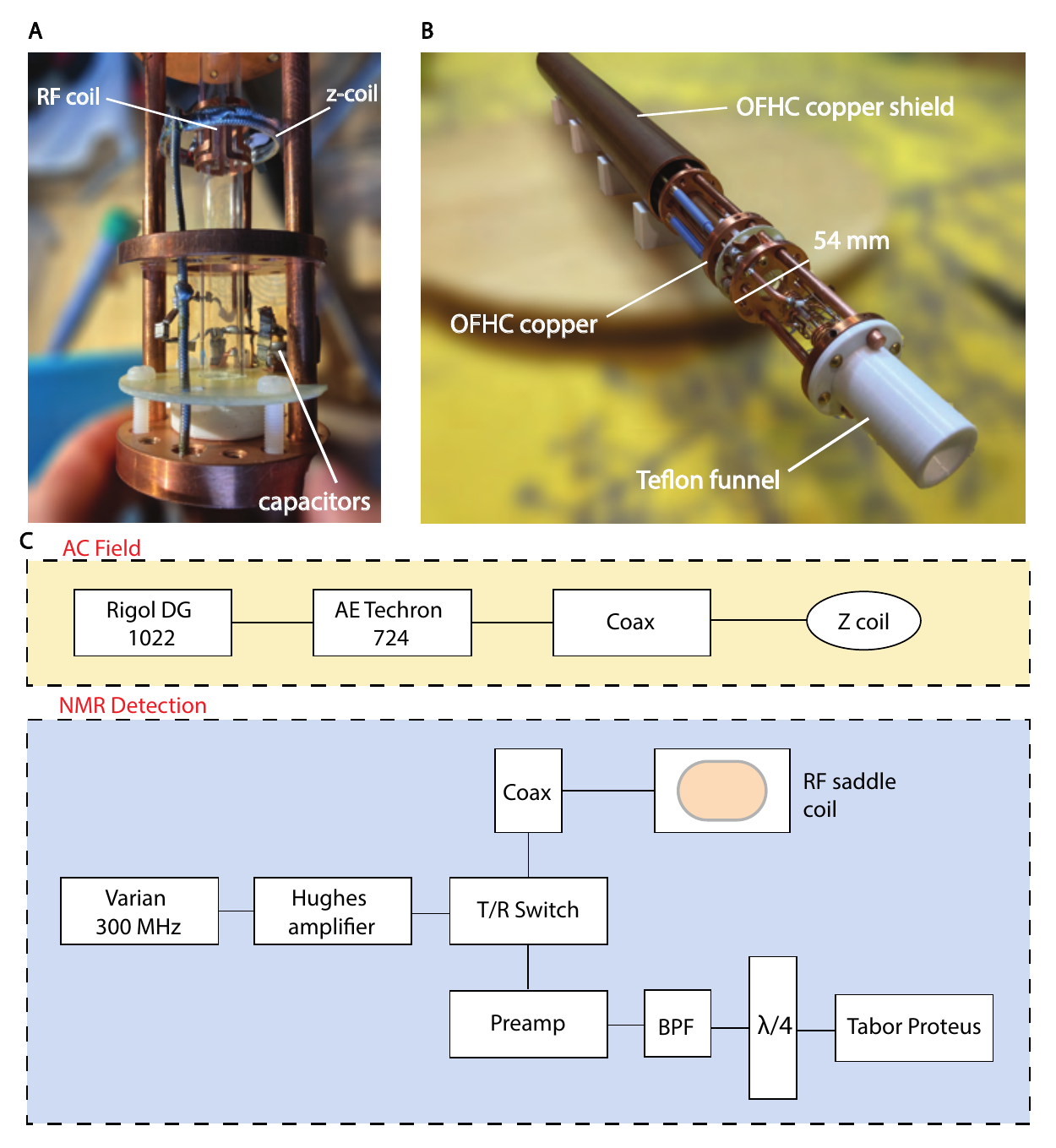}}
  \caption{\T{NMR Probe} used in experiments. (A) Photograph of internal probe components showing (i) RF coil used for $\Cs$ NMR and (ii) z-coil by which the test AC field is applied. Both coils connect to independent rigid coaxial cables, but share a common ground. The diamond sensor is held under water in a test tube and is shuttled into the center of the RF coil (marked). (B) Zoomed out picture of probe highlighting its oxygen free (OFHC) copper construction, surrounded by a 54 mm OFHC copper shield (marked). (C) \I{Circuit. } Panel displays circuit employed in the experiments. The orange region shows circuit used for AC field application. The blue region shows NMR excitation and detection circuit.}
\zfl{probe}
\end{figure}

\begin{figure*}[t]
  \centering
  {\includegraphics[width=0.95\textwidth]{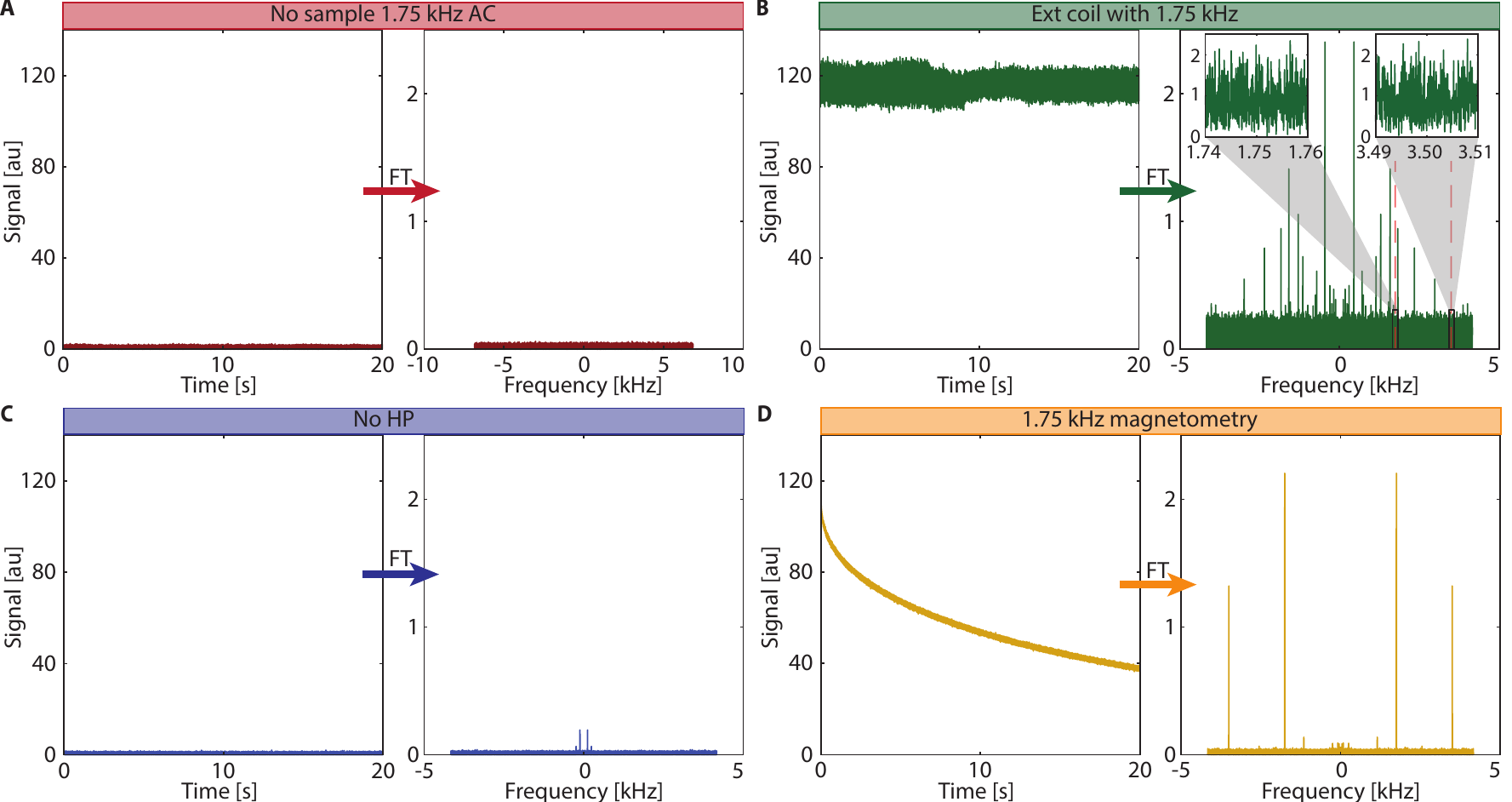}}
  \caption{\T{Experiments testing cross-coupling} between NMR detection and AC field coils. Panels show single-shot pulsed spin lock data and Fourier transforms of $S_o$ (as shown in \zfr{fig2}C of the main paper). We apply a 1.75 kHz AC field in each experiment. (A) Data with no sample in probe. Signal obtained is just noise and Fourier transform shows no signatures of the AC field. Small peaks arise due to noise from probe. (B) Data with simulated field at 75 MHz applied by an additional x-coil, which we refer to as the external coil, with no sample in probe. This simulates precession of $\Cs$ nuclei for the NMR receiver. Time domain data shows a flat non-decaying signal. Fourier transform shows no components at the applied AC frequency. Dashed orange lines indicate the frequencies at which we would expect to see the first and second harmonics.  \I{Inset:} Zoomed into these regions for clarity. Signals at 1.75 kHz and 3.5 kHz are absent. (C) Data with diamond thermally polarized after 10 s in the magnet. Weak SNR without hyperpolarization makes AC field components indiscernible. (D) Reference data performed with hyperpolarized sample (corresponding to \zfr{fig5} of the main paper) showing time domain oscillations and Fourier peaks at 1.75 kHz.}
\zfl{cross-expt}
\end{figure*}

%\part{Supplement}
\section{Scaling of primary and secondary harmonic intensities} 
\zsl{sec1}

In this section, we present extended data on how \zfr{fig5}H was measured, demonstrating how relative scaling of signal intensities between the first and second harmonics changes with respect to pulse width $t_p$. We perform in \zfr{scale} a series of experiments with conditions similar to \zfr{fig6}, conducting an AC field chirps in a 1-4 kHz window ($\xD\mB{=}$3 kHz) in 20 s. Here, we varied pulse width while fixing the  acquisition windows at $\tacq{=}32\;\mu$s. Dead times between pulses and acquisition were also kept constant.  Following \zr{res} then,  the resonance condition shifts with $t_p$.  This is evident in \zfr{scale}A, where we plot the zoomed response of the first harmonic (red) and second harmonic (blue dashed) respectively, while normalizing the peak of the second harmonic profile. We observe that the relative strength of the second harmonic intensity increases with pulse duty cycle. 

From \zfr{scale}A, the resonance frequency is obtained from the cusp of the primary harmonic response (pink line), and also corresponds to half the frequency at which the second harmonic response (dashed blue line) is maximum. \zfr{scale}B (same as \zfr{fig5}H in the main paper) plots this precise measurement of the resonance frequency $\fres$ for differing pulse sequence parameters. Data points show a good fit to the expected dependence of the resonance frequency (\zr{res}),  wherein we extract $\xt{=}\pi/2$ for $t_p{=}36\mu$s. \zfr{scale}C elucidates the extracted ratio of the second to first harmonic intensities. The data  indicates that the second harmonic response is related to finite pulse widths employed and will be absent in the limit of $\delta$-pulses. 

\section{NMR Probe}
\zsl{sec2}
We now provide details of the NMR probe employed in the high field magnetometry experiments at 7 T. For this, we designed and built an NMR probe (see \zfr{probe}A-B) that:

\I{(i)} is capable of high fidelity $\Cs$ inductive detection at 75 MHz, 

\I{(ii)} with high RF homogeneity, allows $\sim$275k pulses to be applied to the nuclear spins at high power ($\sim$30 W) and high duty cycle (${\sim}$50\%, applied every 73 $\mu$s) (\zfr{fig2}A), and 

\I{(iii)} permits the application of a time-varying (AC) magnetic field simultaneous with the pulse sequence (\zfr{fig2}A). 

This is accomplished with a combination of RF and z-coils as shown in \zfr{probe}A. The RF coil employed for $\Cs$ readout is in a saddle geometry and is laser-cut out of OFHC copper with 3 turns and a coil height of 1cm. We measure a Q-factor of 50 at 75 MHz and a sample filling factor of $\sim$0.3. The z-coil employed to apply the AC field is a loop of a 2-turn coaxial cable that minimizes electric fields (\zfr{probe}A). 

\zfr{probe}C shows the circuits employed in these experiments. High SNR RF detection of the $\Cs$ nuclear precession is obtained via a quarter wave line and bandpass filter combination following a transmit/receive switch, and the signal is digitized by a high-speed arbitrary waveform generator (Tabor Proteus). The high data acquisition rate of the device (1 GS/s) allows for high-fidelity sampling of the $\Cs$ induction signal between the pulses. We refer the reader to Sec.~\zsr{sec4} for more details on data processing. The AC field is applied with a Rigol DG1022 signal generator, weakly amplified by an AE Techron 724 amplifier to provide sufficient current swing. This is useful in the experiments shown in \zfr{fig6} and \zfr{fig7} of the main paper, wherein frequency sweeps are employed to determine the frequency response of the sensor. 

\section{Experiments probing coil cross-coupling}
\zsl{sec3}
We performed a series of experiments to ensure that the applied z-coil signals are not picked up by the RF coil. This would eliminate the possibility that the oscillatory $\Cs$ dynamics in \zfr{fig5} arise from coil cross-coupling.  We note at the outset that such pickup is expected to be negligible because:

\I{i.}	The RF and z-coils are orthogonal to ${<}$1$^{\circ}$ and have little mutual inductance coupling, and

\I{ii.}	The applied AC fields are in the 10 Hz-10 kHz range, far outside the detection range of the NMR RF circuit (tuned to 75 MHz$\pm$30 kHz). These applied AC signals are strongly suppressed by the quarter wave line and bandpass filters in the circuit, which deliver a ${>}$80 dB suppression.

Simple experiments bear out this intuition (\zfr{cross-expt}). In these measurements, we applied a 1.75 kHz AC field of 100 mVpp amplitude with the Rigol signal generator and found the corresponding signal in the output under the following conditions:

\I{i.}	In \zfr{cross-expt}A, we removed the sample completely from the probe. The pulses (as in \zfr{fig2}A of the main paper) were then applied and the resulting data was processed by the data handling pipeline shown in \zfr{fig2}. We only observed noise in these measurements (\zfr{cross-expt}A) and a Fourier transform (right panel) did not reveal any signals corresponding to the applied AC field; instead, we only observed noise. 

\I{ii.}	In \zfr{cross-expt}B, we applied a test signal at 75 MHz with an additional x-coil, referred to here as the "external coil", in order to mimic a spin precession signal from the $\Cs$ nuclei. The sample was still absent from the probe in these measurements. As expected, the experimental data reveals a flat non-decaying signal, and a Fourier transform contains a multitude of peaks due to noise. However there are no peaks at 1.75 kHz and 3.5 kHz, the expected positions of the first and second harmonics of the applied AC field.  These are marked by the orange dashed lines, and the insets in \zfr{cross-expt}B show a zoom into these regions, showing only noise. 

\I{iii.}	In \zfr{cross-expt}C, we performed an experiment with the diamond sample thermally polarized for 10 s with no hyperpolarization. Once again, we only observed noise with no signature of the applied AC field.

\I{iv.}	In \zfr{cross-expt}D, the diamond sample was hyperpolarized, resulting in Fourier peaks at 1.75 kHz and 3.5 kHz characteristic of the applied AC field. 

Therefore, we conclude that the AC field harmonics are obtained as a combined action of the spin-lock pulse sequence with the AC field on the hyperpolarized $\Cs$ nuclei. This conclusion is strengthened by the fact that the effective frequency response of the $\Cs$ sensor is not constant, and depends on the exact pulse spacing (see \zfr{fig7} of the main paper) resulting in a sharp response near the resonance condition. 

\section{Data Processing}
\zsl{sec4}
The data processing pipeline employed in this manuscript follows a similar approach as described in Ref. ~\cite{Beatrez21}.  Here, we highlight salient features and emphasize the differences. The NMR signal is sampled continuously in $t_{\R{acq}}$ windows between the pulses (\zfr{fig2}A), at a sampling rate of 1GS/s via the Tabor Proteus. In typical experiments as in \zfr{fig2}A, the pulses are spaced apart by $\qt{=}73\;\mu$s and the acquisition windows are $t_{\R{acq}}{=}32\;\mu$s. The spin precession is heterodyned to 20 MHz which is the oscillation frequency sampled in the measurements (as shown in \zfr{fig2}C). For each acquisition window, we take a Fourier transform and extract the 20MHz peak as in \zfr{fig2} of the main paper. This corresponds to the application of a digital bandpass filter with a linewidth of $t_{\R{acq}}^{-1}{\app} 31.2$kHz. For the 20 s long acquisition periods employed in the paper and a pulse spacing of 73 $\mu$s, we have $\sim$275k data collection windows.

In the experimental data, we focus on two complementary aspects (highlighted in \zfr{fig3} of the main paper): 
\benum[noitemsep,topsep=0pt,label=(\roman*)]
\item The decay of the pulsed spin-lock signal as the AC field approaches the resonance condition. 
\item Oscillations riding on the pulsed spin-lock signal, which carry the imprint of the AC field applied to the nuclei. 
\eenum
To isolate the decay \I{(i)}, we smooth the pulsed spin-lock data over an interval of 73ms. This corresponds
to a 13.7 Hz digital low-pass filter that suppresses the oscillations. Subtracting the smoothed data from the raw data curve reveals the oscillations \I{(ii)}. 

We emphasize that the $\Cs$ oscillatory dynamics manifest as an amplitude modulation and not from the frequency shift of the spin precession outside the detection window. For example, \zfr{fig2}B-D of the main paper shows the raw data obtained between the pulses and their corresponding Fourier transforms for the signal from an AC frequency of 2 kHz and a voltage intensity of 100 mVpp. Evidently, the AC field intensity is so low that it does not cause a shift in the Fourier transform peaks which remain at 20 MHz. To appreciably shift the frequency here, the AC field would have to be at least 30 G in strength — the fields we employ are two orders of magnitude lower. There is, instead, an amplitude variation between the FT peaks which is plotted in \zfr{fig2}D of the main paper (oscillations which imprint the amplitude and frequency of the applied AC field). 

\begin{figure}[t]
  \centering
  {\includegraphics[width=0.5\textwidth]{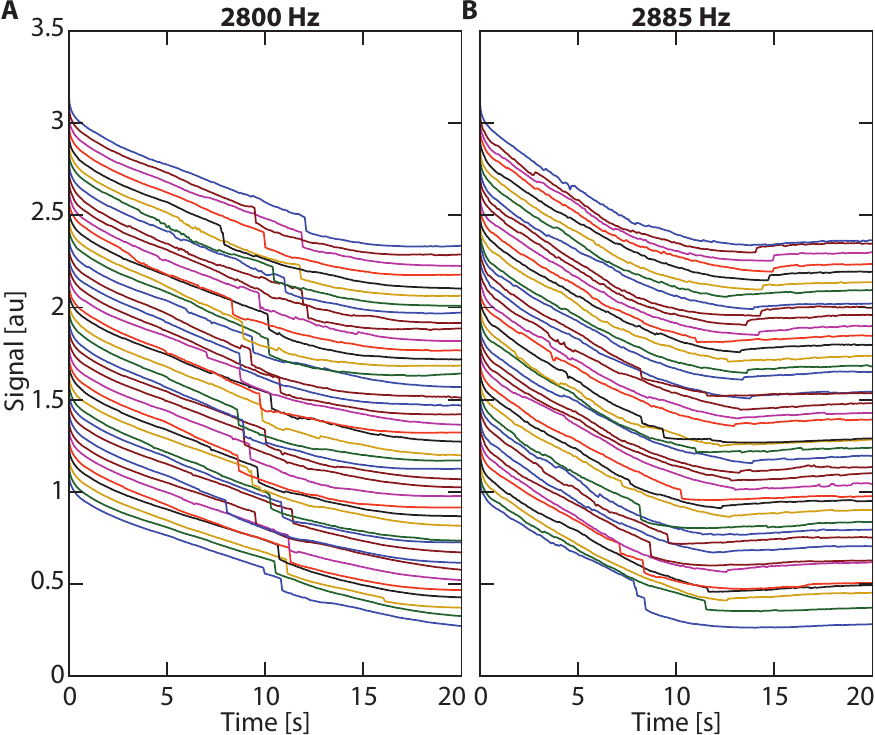}}
  \caption{\T{Signatures of chaos in signal decays}. In decay profiles obtained at (A) $\fac{=}2.8$ kHz, and (B) $\fac{=} 2.885$ kHz, several sudden jumps are observed. Both experiments are repeated 43 times consecutively and plotted with a vertical offset. Location of the jumps differ significantly from one another implying that the behavior is unpredictable and possibly chaotic.}
\zfl{chaos}
\end{figure}

\section{Signatures of chaos in signal decays}
\label{section:chaos}
\zsl{sec5}
Interestingly, we observe signatures resembling chaotic dynamics in the spin-lock profiles as one approaches the resonance condition. This is evident in \zfr{fig4}E-F of the main paper. We postpone a more detailed discussion of the origin of this behavior to a separate manuscript, but present here (see \zfr{chaos}) supplementary data to \zfr{fig4}E-F of the main paper. \zfr{chaos} shows data corresponding to 43 concurrent runs of the experiment, vertically offset for clarity.  Here $\fac{=}2.8$ kHz, and $\fac{=} 2.885$ kHz respectively, close to, and at resonance ($\fres{=}2.885$ kHz) respectively. We observe sharp jumps in the decay profiles; these punctuate the longer periods where the decay is slow. The extent of this possibly chaotic dynamics increases as one approaches resonance and falls sharply away from it (\zfr{fig4}E-F). We also note that we only observe them at high nuclear Rabi frequencies.

\section{Average Hamiltonian Analysis}

\zsl{sec6}
Here we present a more detailed average Hamiltonian analysis for the operation of the magnetometry sequence. For the majority of this section, unless otherwise explicitly stated, we treat the system in the lab frame. The total Hamiltonian of the system is $\mH{=}\mH_{\R{Z}}+\mH_{\R{dd}}+\mH_{\R{AC}}$ where $\mH_{\R{Z}}{=}\wl I_z$ is the Zeeman Hamiltonian, $\mHdd{=}\sum_{k<\ell}b_{k\ell}(3I_{kz}I_{\ell z}-\overrightarrow{I_k}\cdot \overrightarrow{I_\ell})$ corresponds to the dipolar interaction, $\mH_{\R{AC}}{=}\gamma_n \Bac\cos(2\pi \fac t+\phi_0)I_z$ is applied field to be sensed, $\wl=\xg_n B_0$ is the Larmor frequency,  and $\Bac$, $\wac$, and $\phi_0$ are the applied AC field amplitude, frequency, and phase respectively.

We will simplify the dipolar Hamiltonian, separate from the rest of the terms, using average Hamiltonian theory (AHT) by only keeping the leading-order term. These assumptions are reasonable under the condition $\zeta{=}2\pi J\qt{\ll} 1$.  For each pulse period, we will treat the system in the toggling frame defined by the pulses up to that point. For the period after the $j^{\R{th}}$ pulse, the toggling frame transformation is given by
\bea
\mH_\R{dd}^{(j)} = \exp\big(-ij\xt I_x\big) \mH_{\R{dd}}  \exp\big(ij\xt I_x\big),
\zl{toggle}
\eea
Consider the case when $\xt{=}\frac{\pi}{2}$. In this case, the toggling frame wraps to the original frame after a period of $4\tau$.  For a multiple period of $4\tau$ then,  the system dynamics is captured by the average Hamiltonian, 
\bea
\mH_\R{dd}^{(0)}= \sum \mH_\R{dd}^{(j)} = \sum_{j<k}b_{kl}\bigg(\frac{3}{2}(I_{jz}I_{kz}+I_{jy}I_{ky})-\vec{I_j}\cdot\vec{I_k}\bigg).
\eea
Higher order AHT terms are evaluated in detail in Ref. ~\cite{Ajoy20DD}. The initial state $\rho(0){=}I_{x}$ is protected against dipolar coupling because $[\mH_\R{dd}^{(0)},\rho(0)]{=}0$. Since the state is protected over an average, we will neglect the dipolar term in all of the following sections.

We now treat the external field Hamiltonian, $H_{\R{AC}}$ using the same method, once again assuming $\xt{=}\pi/2$.  A DC field, whose lab frame Hamiltonian is given by $H_{\R{AC}}{=}\gamma_nBI_z$, will be rotated by $\pi/2$ by each toggling frame transformation such that over a four pulse period, the Hamiltonian will average to 0. It is possible to map the toggling frame Hamiltonians for each pulse period on a phasor in order to see this more clearly (\zfr{fig4}C). For this case, the toggling frame Hamiltonians will cover the phasor symmetrically such that the vectoral sum on the phasor will be 0. This also works for the more general case $\xt{=}\frac{2\pi k}{n}$ for integers $k$ and $n$, as the points will trace a regular polygon on the phasor, which will still vanish when summed.

However, for a resonant AC field (assumed to be a square wave for simplicity) with frequency $f_{\R{AC}}{=}1/4\tau$, the average Hamiltonian is a linear combination of $I_y$ and $I_z$ (for the special cases where the AC field phase is $\phi{=}-\frac{\pi}{6}$ or $\frac{5\pi}{6}$, the $I_z$ terms cancel out but there is a net $I_y$ term). For any AC field with $f_{\R{AC}}{\neq} f_{\R{res}}$, the external field Hamiltonian will average to zero, albeit over a longer timescale.

\begin{figure}[t]
  \centering
  {\includegraphics[width=0.5\textwidth]{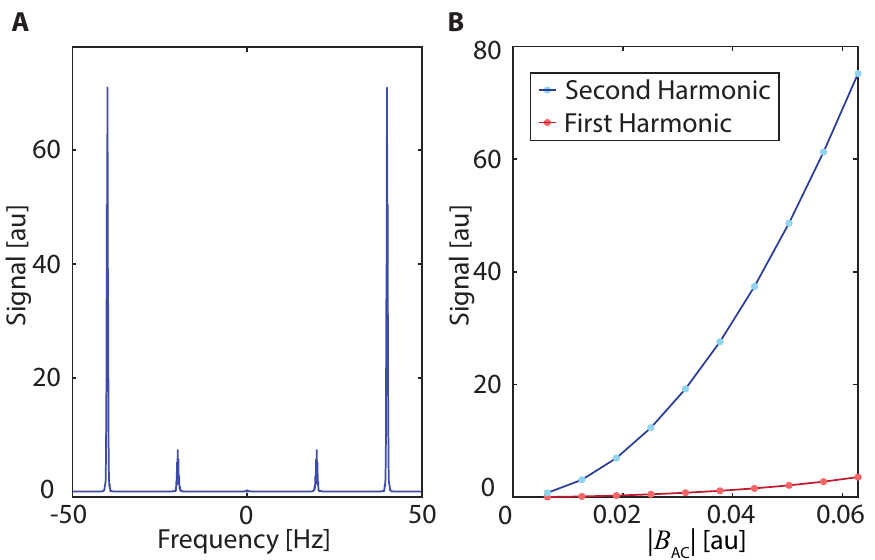}}
  \caption{\T{Simulation of the signal} in the resonant case $\fac{=}\fres{=}20$Hz, following the theoretical expression derived in the main paper. (A) Left panel shows the magnitude of the Fourier transform of the oscillatory component $S_o$ of the signal. It reveals two harmonics at $\fac$ and $2\fac$ respectively. (B) Right panel shows the scaling of the magnitudes of the two harmonics, and shows good qualitative agreement with the experimental results in \zfr{fig5}H.   
}
\zfl{figS5}
\end{figure}

\section{Spin evolution for off-resonance fields}
\zsl{off-res}

In the main paper, we had elucidated theory of the experiment as a ``rotating-frame'' NMR analogue. This calculation was provided for the resonant case, i.e. $\fac{=}\fres.$ \zfr{figS5} shows a simulation of the signal $
S(t) = \lsb \cos^2(\xg_n \Bac t) + \sin^2(\xg_n \Bac t)\sin^2(2\pi \fac t)\rsb^{1/2}
$, obtained for a representative example of $\fac{=}$20Hz. Plotted is the magnitude of the Fourier transform of the oscillatory signal $S_o$ similar to Fig. 3C and \zfr{fig5} of the main paper. The Fourier transform in \zfr{figS5}A reveals the precedence of two harmonics (primary and secondary). \zfr{figS5}B plots the magnitudes of both harmonics, which is in good qualitative agreement with the data in \zfr{fig5}H.

In this section, we extend this calculation to the off-resonant case, i.e. for arbitrary applied frequency $\fac$.  Here we neglect the effect of dipolar interaction driven evolution, assuming it is suppressed via the analysis above. Consider first that the rotating frame Hamiltonian (at $\xo_L$) can be written of the form,
\beq
\mH = \xO I_x + \xg_n\Bac\cos(2\pi\fac t)I_z
\eeq
In a second rotating frame at $2\pi\fac$ and assuming a rotating wave approximation $\xg_n|\Bac|{\ll} \fac$, one can write down the state evolution as,
\beq
\xr'_R(t) = \cos\xa\cos(Qt)I_x + \cos\xa\sin(Qt)I_y + \sin\xa I_z,
\eeq
where $\tan\xa{=} (\xO-2\pi\fac)/(\xg\Bac)$, and $Q{=}\cos\xt\lsb (\xO-2\pi\fac)\tan\xa + \fr{1}{2}\xg\Bac\rsb$. In the original rotating frame, the state then has the form,
\beq
\xr_R(t) = \lb \cos^2\xa\cos (Qt) + \sin^2\xa\rb I_x + \cos\xa\sin (Qt)I_y.
\eeq
Finally returning to the lab frame where the measurement is carried out,  yields the state evolution,
\bea
\xr(t) &=&  \lb \cos^2\xa\cos (Qt) + \sin^2\xa\rb I_x + \cos\xa\sin (Qt)\cos(2\pi\fac t) I_y \non\\
&+&  \cos\xa\sin (Qt)\sin(2\pi\fac t) I_z.
\eea
Ultimately, since we measure $S=[\expec{I_x}^2 + \expec{I_y}^2]^{1/2}$,  there is an oscillation of the form,
\beq
S = \lsb\lb \cos^2\xa\cos (Qt) + \sin^2\xa\rb^2 + \cos^2\xa\sin^2 (Qt)\cos^2(2\pi\fac t)\rsb^{1/2}
\eeq
that carries first and second harmonics of $\fac$,  similar to \zfr{figS5}.

\begin{figure}[t]
  \centering
  {\includegraphics[width=0.5\textwidth]{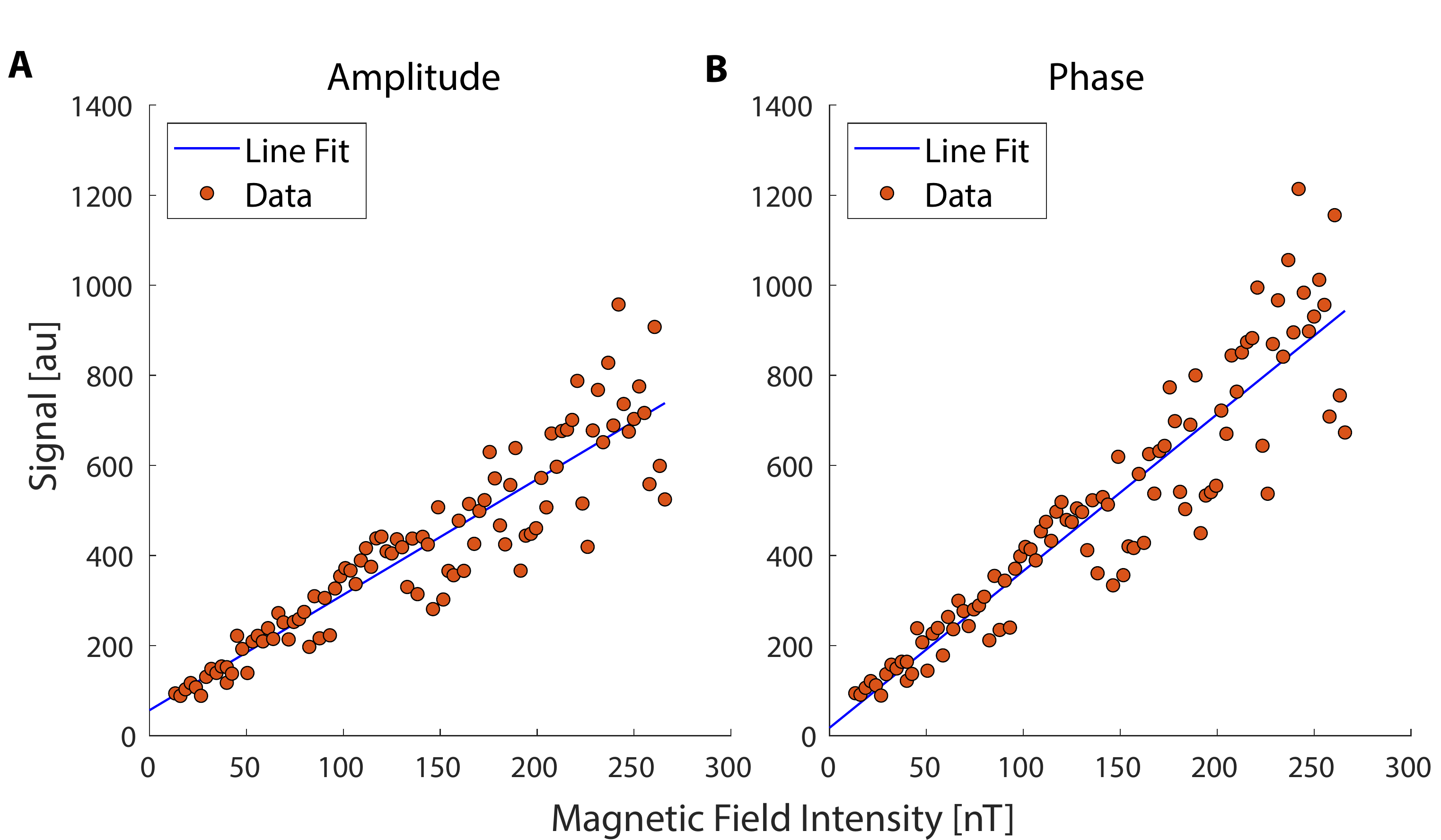}}
  \caption{\T{Sensitivity analysis. } Points show the measured strength of the first harmonic of the oscillatory signal  for varying strengths of the AC field $|\Bac|$ obtained via (A) the magnitude  and (B) phase of the FT signal as in \zfr{fig2} of the main paper. Data here is carried out for $t{=}34$s of signal acquisition. For each case, a linear extrapolation of the data towards zero allows us to estimate the sensitivity.}
\zfl{figS5_1}
\end{figure}

\begin{figure*}[t]
  \centering
  {\includegraphics[width=0.95\textwidth]{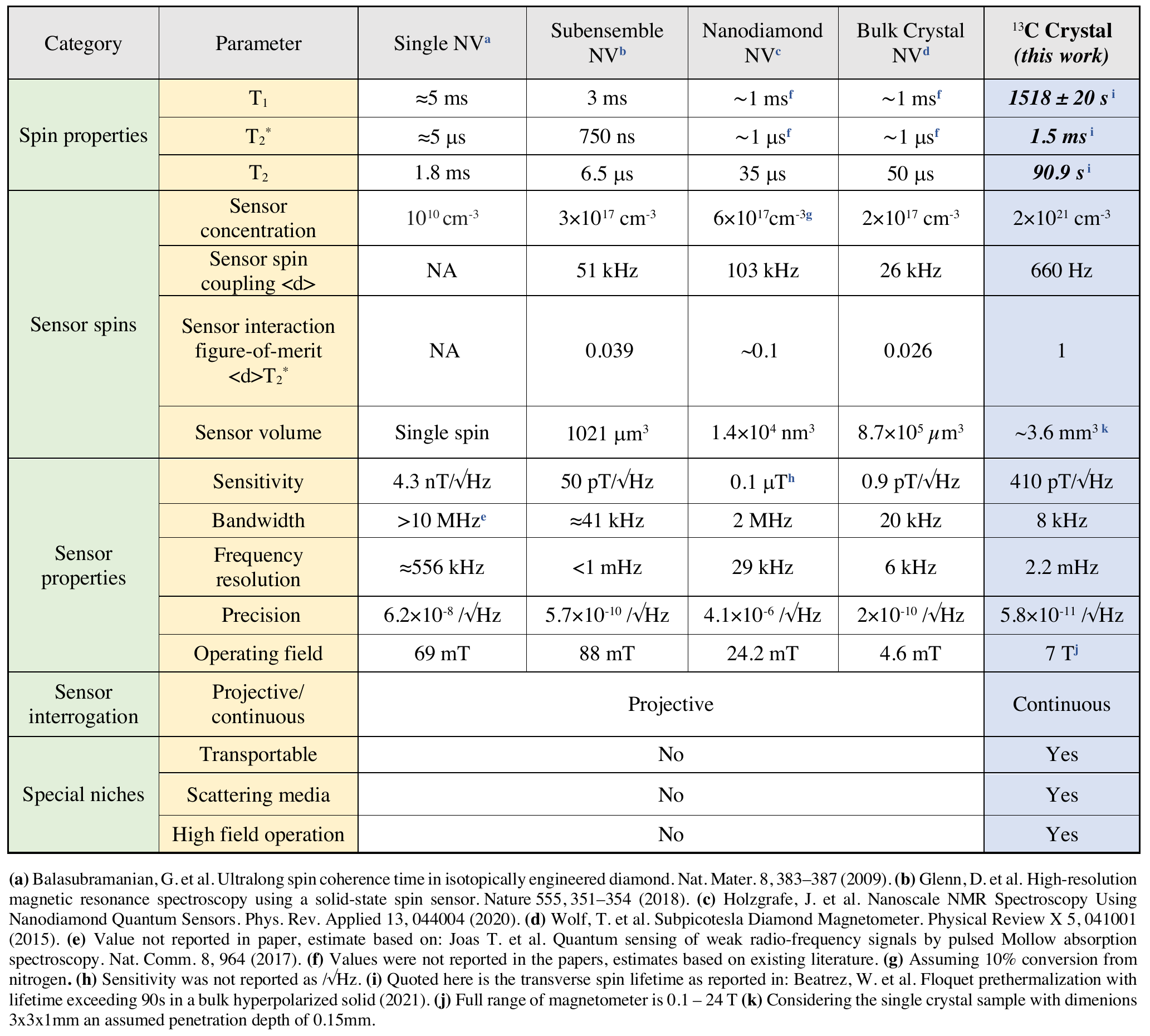}}
  \caption{\T{Comparison between NV center magnetometers and $\Cs$ nuclear sensors} considering representative literature for different sensor sizes. $\Cs$ sensors have complementary properties to NV sensors and are attractive in some operating niches.}
\zfl{table}
\end{figure*}

\section{Simulation of Sequence Performance}

\subsection{Dipolar Network Simulation}
\zsl{sec7a}
For the result in \zfr{fig4}B, we numerically simulate a sequence operation assuming a Hamiltonian of the form, $\mH =  k_{\R{dd}} \mHdd+k_z\mH_z$ where $\mHdd$ and $\mH_z$ are both normalized such that, $\|\mHdd\|=1$, where $\|.\|$ refers to the Frobenius norm.

We assume application of a square-modulated AC magnetic field at frequency $\fac$. We consider  the average results from 10 random configurations of a $N{=}5$ spin $\Cs$ nuclear network, and simulate dynamics under the full spin Hamiltonian $\mH$. Let $t_j$ denote “switching events”, i.e. time instants where either a pulse is applied, or there is a sign switch in the AC field. The propagator then evaluates to, $U{=}\prod_j U_j$, where $U_p{=}\exp \left[i\left(\kdd\mHdd+k_{z} H_{z} s_{j}\right)\xD\qt_j\right]$, and $\xD\qt_j{=}(\qt_{j+1}-\qt_j)$, and $s_j$ is a sign term with $s_1{=}1$. We have \I{(i)} in case of a pulse event, $U_j{=}U_pe^{i\xt I_x}$, with $s_{j+1}{=}s_j$; \I{(ii)} upon a field sign switch $U_j{=}U_p$, with $s_{j+1}{=}-s_j$; and \I{(iii)} when both occur simultaneously, $U_j{=}U_pe^{i\xt I_x}$, with $s_{j+1}{=}-s_j$. The sequence fidelity is then evaluated  via the survival probability in the $\xhat\tm\yhat$ plane.

In order to obtain the linewidth scaling with respect to the number of pulses as shown in \zfr{fig4}G, we run this simulation for different numbers of pulses (all multiples of 4) and the resulting frequency response is fit to a gaussian profile peaking around the resonance frequency. The linewidth is then extracted as the FWHM of the fit and plotted as a function of the number of pulses (L). This process is then repeated for different dipolar coupling strengths ($k_{\R{dd}}$).

In order to compare this numerical result with zeroth-order AHT, we run a similar simulation for AHT where instead of computing the propagator for each time event, we record the toggling frame "angle" for the time block between that event and the next. A pulse rotates the toggling frame by $\xt$ and a sign flip of the square wave rotates it by $\pi$. We then compute the average Hamiltonian for $\mH_z$ to be $\Hzav=\sum_j\exp(i\xt_j)\Delta t_j$ where $j$ is the index for a given time block, $\xt_j$ is the toggling frame angle for that time block, and $\Delta t_j$ is the length of that time block. The real part of the resulting complex number corresponds to the coefficient of the $I_z$ term and the imaginary part, the $-I_y$ term. The dipolar term is treated similarly. The effect of a toggling frame rotation by $\xt$ on the dipolar Hamiltonian is given by,

\bea
\mH_{\R{dd}}^{(\xt)}&=&\sum_{j<k}\cos^2(\xt)I_{jz}I_{kz}+\cos(\xt)\sin(\xt)(I_{jy}I_{kz}+I_{jz}I_{ky})\non\\
&&+\sin^2(\xt)I_{jy}I_{ky}.
\eea

The time propagator then becomes $U=\exp(-i(\Hddav+\Hzav)t)$ where $t=\sum_jt_j$ is the total time run by the simulation and the fidelity is once again evaluated.  Both simulations are then run for multiple AC frequencies in order to obtain the results for \zfr{fig4}B.

\begin{figure*}[t]
  \centering
  {\includegraphics[width=1\textwidth]{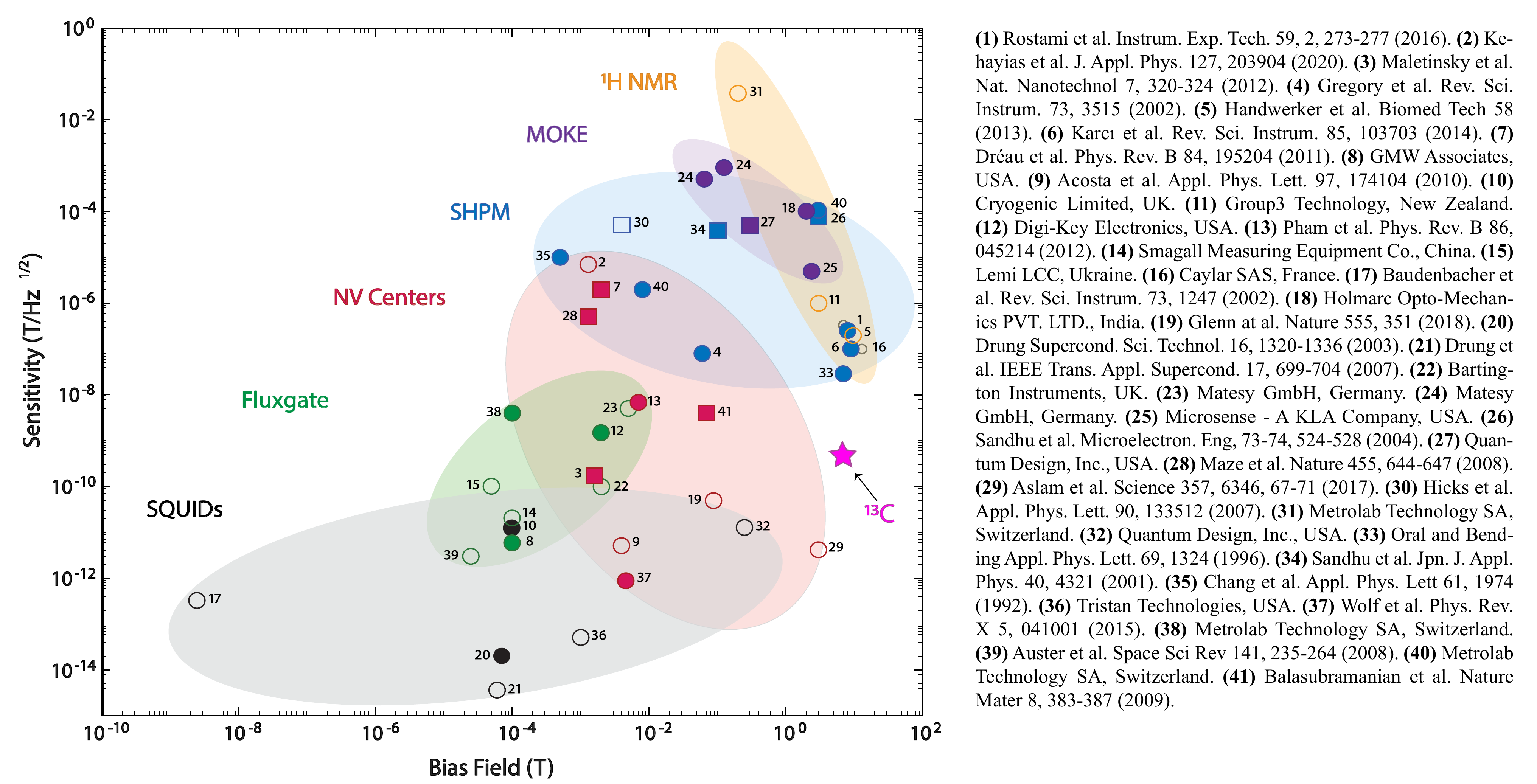}}
  \caption{\T{Magnetometer landscape} focusing on a comparison with respect to sensitivity and bias field of operation. Points quote values from literature listed in footnote (right). Micro/nano sensors and macro sensors are depicted as squares and circles respectively. Filled points represent AC magnetic field sensors and outlined points represent DC sensors. $\Cs$ magnetometers (depicted as a star) occupy a niche for high sensitivity magnetometry at high bias field.}
\zfl{landscape}
\end{figure*}

\subsection{Delta Pulse Simulations}
\zsl{sec7b}

The general spin trajectory under a $\pi/2$ pulse train as depicted in \zfr{fig1}B is obtained by considering a single spin system initialized in the $\rho(0)=\cos(\frac{\pi}{6})I_x+\sin(\frac{\pi}{6})I_y$ state, a $\frac{\pi}{6}$ deviation from the $\xhat$-axis and on the $\xhat\tm\yhat$ plane. The spin is under the previously mentioned pulse sequence and an external AC field resonant with that sequence ($f_{\R{AC}}{=}f_{\R{res}}{=}\frac{1}{4\tau}$) along the $\zhat$-direction with an intensity of 16 nT. Our strategy is to numerically solve the classical Bloch equations for this system in the rotating frame. We achieve that by doing a finite element analysis with a time step such that each pulse period has 2000 points ($\sim$85 ns) and rotating the state along the $\zhat$-axis by $2\pi\gamma_nB(t)\Delta t$ where $B(t)$ is the magnetic field intensity at time $t$ and $\Delta t$ is the timestep. If a pulse happens after a timestep, the state is rotated by $\pi/2$ along the $\xhat$-axis. The simulation is run for 100 cycles of the AC field period and plotted on a Bloch sphere in order to obtain the graph in \zfr{fig1}B.

\section{Estimation of sensitivity}
\zsl{sens}
To estimate the sensitivity of the $\Cs$ sensor, we perform a careful calibration experiment with an AC fields of known intensity and compare against the strength of the signal response for the first harmonic of the obtained oscillatory signal. In particular, in these experiments, we apply an AC field via an calibrated current input through z-coil in \zfr{probe}. The current is measured via the voltage drop through a series resistor, and the field is estimated using Biot-Savart law. 

We then measure the strength of the oscillatory signal for different values of $|\Bac|$, measuring them via the magnitude and phase of the FT signal in \zfr{fig2} of the main paper.  This is shown in \zfr{figS5_1}.  Measurements here are carried out in a single shot for $t{=}34$s. The phase signal concentrates all the intensity in the first harmonic and hence carries higher sensitivity. However, it needs to unwrapped to extract only the phase of the spins in the rotating frame (removing the trivial phase accrued during the $t_p$ pulse periods). We will detail a numerical procedure for this unwrapping in a forthcoming paper. In each case, we estimate sensitivity by fitting the obtained signal to a linear fit.  Interpolating the fit allows us to discern the minimum magnetic field that would produce a signal change on the order of the noise level. Through this, we estimate a smallest measurable signal in a single-shot as $130\pm 22$ pT and $70\pm 15$ pT respectively, yielding a sensitivity of $760\pm 127$ pT$/\sq{\R{Hz}}$ and $410\pm90$ pT$/\sq{\R{Hz}}$ respectively at 95\% CI. These are the results quoted in the main paper.

\section{Comparison between NV and $\Cs$ magnetometers}
\zsl{sec10}
Here we contrast the key features of $\Cs$ nuclear spins and NV centers focused on applications in magnetometry. We include in this comparison four representative papers from the literature for different regimes of NV center quantum sensors - \I{(i)} single NV electrons well-isolated in the lattice~\cite{Balasubramanian09}, \I{(ii)} sub-ensembles of NV centers (occupying ${<}100\;\mu m^3$) in single crystal samples~\cite{Glenn18}, \I{(iii)} ensembles of NVs in microdiamond~\cite{Holzgrafe20}, and \I{(iv)} dense NV centers in bulk single crystal samples~\cite{Wolf15}. For each, we elucidate values of key spin parameters, and in the last column in \zfr{table} compare them to corresponding properties of $\Cs$ nuclear spins for the single crystal sample used in this work (same as that employed in Ref.~\cite{Beatrez21}). This table (\zfr{table}) shows the complementary properties of NV electrons and $\Cs$ nuclei for sensing. To make the comparison clearer, we focus on five major aspects as detailed below —

\I{i.	Spin properties:} focusing on the respective values of $T_2^*$, $T_2$, and $T_1$. For $T_2$, we quote the values obtained pulsed driving (e.g. DD) that is relevant for quantum sensing. For the $\Cs$ case, we quote the rotating frame lifetime value $T_2’$ under pulsed spin locking~\cite{Beatrez21}. This sets the total interrogation time that the $\Cs$ sensors can permit. We qualify that this does not correspond to the total time over which the $\Cs$ sensors can continue to accrue phase; which instead depends on an interplay between ${T_2^{\ast}}$ and $\fac$, and is more difficult to estimate. We do emphasize however that for the $\Cs$ case, the extension in the $T_2'$ value, ${T_2'}/{T_2^{\ast}}$ (${\app}60,000$ in \cite{Beatrez21}), is considerably larger than a naive scaling from the corresponding electron spin values by just the ratio of gyromagnetic ratios.

\I{ii.	Independent or coupled sensors:} In the Table (\zfr{table}), we estimate the coupling strengths  $\expec{d}$ between the quantum sensors in each of the samples considered. Even at natural abundance, the $\Cs$ sensor concentration at least four orders of magnitude denser than NV center sensors. Evidently then, NV center quantum sensors largely operate in the limit of uncoupled (independent) sensors, wherein their $T_2^{\ast}$ times are dominated by interactions with other spins as opposed to inter-sensor interactions.  In contrast, $\Cs$ sensors operate in the limit where interspin couplings, scaling with enrichment as $\eta^{1/2}$, dominate the $T_2^{\ast}$ FID times. This can be seen by comparing the product $\expec{d}T_2^{\ast}$.  Our pulsed spin-lock scheme is able to mitigate the effect of these couplings, allows interrogation up to $T_2’$.

\I{iii.	Sensor properties for AC magnetometry:} focusing on sensitivity, bandwidth, spectral resolution, precision, and operating field.  We note that while our experiments were carried out on a diamond crystal that was uniformly illuminated, we estimate a penetration depth ${<}$0.15mm~\cite{Sarkar21,Acosta09}.  Sensitivity refers to the smallest field that can be reproducibly measured over the bias magnetic field.   Precision here refers to the magnetic field sensitivity by the bias field. In contrast to the lower field operation of the NV center quantum sensors, the $\Cs$ sensor operates at 7 T. We estimate that the full dynamic range reaches can exceed 24 T, presenting the key strength of the approach here.  Assuming a measurement up to 573 s (as demonstrated in Ref. ~\cite{Beatrez21}), a frequency resolution of 2.2 mHz is viable for the $\Cs$ sensor, without reinitialization.

\I{iv.	Sensor interrogation:} focusing on how in each case the sensors are prepared and read out. We note that $\Cs$ readout is not projective and can proceed simultaneously with the magnetometry pulse sequence without reinitialization. Once the spins are polarized, the sensing period can last up to 10 min~\cite{Beatrez21}. NV center sensors must be reinitialized before readout. 

\I{v. Special niches:} Finally, in Table (\zfr{table}), we summarize some special features of the $\Cs$ sensors. This can special niches for the operation of these sensors. First, the long $T_1$ time can potentially allow a separation between of field regions corresponding to $\Cs$ initialization, sensing, and readout. This can enable transportability of the magnetometer between different field regions. Second, RF interrogation allows operation in optically dense or scattering media. Finally, their low magnetogyric ratio allows operation at high magnetic fields.

\section{Comparison to other high-field magnetometers}
\zsl{sec11}
Our $\Cs$ sensors are particularly suited for the high-field magnetometry of time-varying signals. Particularly, their ability to obtain a high frequency resolution (${\sim}2.2$ mHz) over a 8 kHz bandwidth in a wide range of magnetic fields is an important advantage over competing magnetometers.  In this section, we compare $\Cs$ sensors to common magnetometry techniques. We have included SQUID, Fluxgate, Scanning Hall Probe Microscopy (SHPM), Magneto-Optic Kerr Effect (MOKE), and $^{1}$H NMR magnetometers in this comparison because of their ubiquity,  and NV centers because of their close relation to $\Cs$ sensors.

\zfr{landscape} graphically depicts a \I{“landscape”} of these sensor technologies, focusing on their sensitivity and bias field of operation. Sensitivity (y-axis) is defined as the smallest measurable field $\delta B$ over the bias field $B_0$ (x-axis) and is smaller for better magnetic field sensors. Because sensitivity depends, in general, on sensor size,  \zfr{landscape} draws a distinction between nano/microscale (squares) and macroscale (circles) sensors, as well as AC field sensors (filled) and DC field sensors (outlined). Shaded regions show representative values for each technology. 

Hyperpolarized $\Cs$ sensors, as described in this paper, are illustrated by the pink star in \zfr{landscape}. $\Cs$ magnetometers present advantageous with respect to SQUID, Fluxgate, and NV center technologies regarding their ability for AC sensing detection and high field operation. The sensitivity reported in this paper, 410 pT$/\sqrt{\text{Hz}}$,  is better than that of MOKE, SHPM and ${}^{1}$H NMR high-field sensors,  while also allows AC magnetometry with high-frequency resolution.  We expect to further improve sensitivity with technical enhancements of our hyperpolarization and measurement apparatus.

\end{document}